\documentclass[aps,prc,reprint,groupedaddress]{revtex4-1}
\pdfoutput=1
\usepackage{graphicx}

\usepackage{slashed}
\usepackage{amsmath}
\usepackage{amsfonts}
\usepackage{dsfont}
\usepackage{xcolor}
\usepackage{multirow}                            % Multicolumn & multirow
\usepackage{rotating}
\usepackage{array}
\usepackage{colortbl}
\usepackage{latexsym}                            % Load additional symbols
\usepackage[mathscr]{eucal}                      % Euler font symbols
\usepackage{hyperref}
\numberwithin{equation}{section}

\definecolor{airforceblue}{rgb}{0.36, 0.54, 0.66}

\begin{document}

%\preprint{}

\title{Hybrid Stars using the Quark-Meson Coupling and Proper Time NJL Models}


\author{D.~L.~Whittenbury}
\email[]{daniel.whittenbury@adelaide.edu.au}
\author{H.~H.~Matevosyan}
\author{A.~W.~Thomas}
%\thanks{}
\affiliation{CSSM and ARC Centre of Excellence for Particle Physics at the 
Terascale,\\ School of Physical Sciences,
University of Adelaide,
  Adelaide SA 5005, Australia}


\date{\today}

\begin{abstract}
\begin{description}
\item[Background] At high density deconfinement
of hadronic matter may occur leading to quark matter. The immense densities reached in the inner core of massive neutron stars may be sufficient to facilitate the transition.

\item[Purpose] 
To investigate a crossover transition between two phenomenological models which epitomise QCD in two different regimes, while incorporating the influence of quark degrees of freedom in both. 

\item[Method] 
We use the Hartree-Fock quark-meson coupling model and the proper time regularised three flavour NJL model to describe hadronic and quark matter, respectively.  Hybrid equations of state are obtained by interpolating the energy density as a function of total baryonic density and calculating the pressure.

\item[Results] Equations of state for hadronic, quark and hybrid matter and the resulting mass versus radius curves for hybrid stars are shown, as well as other relevant physical quantities such as species fractions and the speed of sound in matter.

\item[Conclusions]  The observations of massive neutron stars can certainly be explained within such a construction. However, the so-called thermodynamic correction arising from an interpolation method can have a considerable impact on the equation of state. The interpolation dependency of and physical meaning behind such corrections require further study.

\end{description}
\end{abstract}

%\keywords{}

\maketitle

\section{Introduction}
\label{sec:intro}

It has long been thought that the densities reached inside the inner core of 
neutron stars may be sufficient to produce a phase transition
from hadronic matter to deconfined quark matter forming hybrid stars. A transition to the chirally restored phase of QCD is also thought to occur at high density. It is unknown if these two transitions coincide and the form they may take.
In understanding these transitions one would ideally like to use QCD directly, but this is currently too difficult. One typically resorts to using two phenomenological models which epitomise the key features of QCD in the two asymptotic regions of its phase diagram\textemdash one in the low density region modelling hadronic matter and the other modelling quark matter at intermediate to high density\textemdash and then construct a phase transition between the two. This means that the dissociation of hadrons does not occur naturally and is dependent on how we construct the transition.

Various models have been used to describe each phase and different constructions in characterising the process of deconfinement have been investigated, such as the Maxwell~\cite{Rosenhauer1991,Klahn:2006iw,Blaschke2008,PhysRevD.81.023009,Blaschke:2010vd,Bonanno2012,Lenzi2012,Logoteta2013,Plumari2013,Benic2014,Klahn2015}
 and Gibbs~\cite{Glendenning:1992vb,Schertler:1999xn,Bentz:2002um,Lawley:2006ps,Carroll:2008sv,Grunfeld2008,Xu2010,Shao2011,Logoteta2012a,Logoteta2012b,Orsaria2013,PhysRevC.90.045801,Miyatsu2015} constructions describing first order transitions, as well as interpolation/percolation constructions where the transition is taken to be of crossover type~\cite{Asakawa1997,Masuda:2012kf,Masuda:2012ed,PhysRevC.90.045801,AlvarezCastillo2014,Kojo2015}. The calculated properties of hybrid stars are considerably influenced by the choice of models and the type of construction used to describe the transition. 
In this work we will investigate transitions of the second or crossover type.
This kind of transition was, for example, recently investigated by Masuda {\it et al}~\cite{Masuda:2012kf,Masuda:2012ed} using several hadronic models and the three momentum regularised NJL model.  They concluded that massive hybrid stars ($M_{\rm NS}\sim 2M_{\odot}$) could be produced using a percolation picture provided the quark matter EoS was stiff enough and the transition occurred at moderately low density ($\rho\sim 3\rho_{0}$), thus providing a possible reconciliation of exotic degrees of freedom with the observations of Demorest {\it et al.}~\cite{Demorest:2010bx} and Antoniadis {\it et al.}~\cite{Antoniadis:2013pzd}. 
We now examine this possibility using the Hartree\textendash Fock QMC model~\cite{PhysRevC.89.065801} and both the proper\textendash time and three momentum regularised versions of the NJL model.

In using these two models, quark degrees of freedom influence both regions, with the latter also exemplifying chiral symmetry breaking. With both models employing quark degrees of freedom it is hoped that they will be more reliable in the transition region. In this region, where hadrons and quarks are expected to coexist, it is likely that the quark substructure of hadrons would play an important role and their interaction with the external quarks strong. The QMC model has the advantage over models which employ point-like descriptions of hadrons by modelling the baryons as MIT bags. 
By incorporating the change of this internal structure in-medium it naturally incorporates many-body forces~\cite{PhysRevLett.93.132502,Guichon:2006er}, which are in fact crucial for nuclear saturation.

Within the QMC model the in-medium changes of the baryon masses are calculated through the bag equations and then parametrised as functions of the scalar field as
\begin{equation}
M^*_B = M_B - w_{\sigma B}g_{\sigma N}\bar{\sigma}
+ \frac{d}{2}\tilde{w}_{\sigma B}\left(g_{\sigma N}\bar{\sigma}\right)^2 \ ,
\label{eq:massparam}
\end{equation}
(where the weightings $w_{\sigma B}$ and $\tilde{w}_{\sigma B}$ simply
allow the use of a unique coupling to nucleons $g_{\sigma N}$).
 Using this parametrisation and a corresponding density dependent coupling,  
  we can solve for the equation of state in the same standard way as the Walecka 
  model~\cite{Serot:1984ey}, that is at the hadronic
 level. In this way the sub-structure of the baryons is entirely contained in the 
 mass parametrisation. We use the parametrisation given in Ref.~\cite{Guichon:2008zz}, which includes the effects of one gluon exchange. 
 
%%%NJL
The Nambu-Jona-Lasinio (NJL) model is an effective model of the strong interaction introduced in 1961~\cite{Nambu:1961tp,Nambu:1961fr}. Originally the model was formulated in terms of nucleons as a local effective interaction inspired by the BCS theory of superconductivity~\cite{Bardeen:1957mv,Bardeen:1957kj} (At the time of the model's conception quarks were yet to be discovered.). Later it was redeveloped in terms of quarks to model low and intermediate energy QCD.  A number of reviews are available on the NJL model and its applications ~\cite{Vogl:1991qt,Klevansky:1992qe,Hatsuda:1994pi,Ripka:1996fw,Ripka:1997zb,Buballa:2003qv}. Its ability to model the breaking of chiral symmetry dynamically by spontaneously generating mass has made it very popular. 

 The NJL model assumes that at low energy scales the gluons acquire a large effective mass and can be integrated out to a good approximation, leaving a local (contact) four Fermi interaction between quarks. 
Upon integrating out gluons the local colour symmetry of QCD is reduced to a global symmetry and confinement is lost. This shortcoming of the NJL model will not be an issue when modelling quark matter at high density, where the quarks will be considered to be deconfined.

Moreover, the NJL model is not renormalizable. However, unrenormalizable theories are still useful and information can be obtained through the process of regularisation, whereby a cut-off is introduced setting the scale of the model. There are many regularisation schemes in common use.  Here we will use Schwinger's proper time regularisation (PTR)~\cite{Klevansky:1992qe,Ripka:1997zb} and make comparisons to the simple, non-covariant three momentum regularisation (TMR).

%%Previous work
In Ref.~\cite{PhysRevC.89.065801} we extended the QMC model by performing a 
Hartree\textendash Fock calculation including the full vertex structure for 
the vector mesons. This extension only alters the exchange contribution,
including not only the Dirac vector term, as was done in
~\cite{RikovskaStone:2006ta}, but also the Pauli tensor term. These
terms were already included within the QMC model by Krein~{\it et
  al.}~\cite{Krein:1998vc} for symmetric nuclear matter and more
recently by Miyatsu {\it et al.}~\cite{Miyatsu:2011bc}. We generalised the work of Krein~{\it et
  al.}~\cite{Krein:1998vc} by evaluating the full exchange terms for all octet baryons and
adding them, as additional contributions, to the energy density. A consequence
of this increased level of sophistication is that, if we insist on using the 
hyperon couplings predicted in the simple QMC model, with no coupling to
the strange quarks, the $\Lambda$ hyperon is no longer bound.

%%This work

The present line of research complements the recent work of Refs.~\cite{Masuda:2012kf,Masuda:2012ed,PhysRevC.90.045801,AlvarezCastillo2014,Kojo2015}, which also considered deconfinement in neutron stars as a crossover rather than a bona-fide phase transition. The novelty of this work is that we incorporate the influence of quark degrees of freedom in the hadronic phase. Moreover, we use a covariantly regularised NJL model for modelling deconfined quark matter. 
This work also complements Ref.~\cite{Miyatsu2015}, which investigated hybrid star matter using a different variation of the Hartree-Fock QMC model and a bag model, where deconfinement was modelled as a first order transition using a Gibbs construction.  The present version of the QMC model  differs from \cite{Miyatsu2015} as we restrict the exchanged mesons to $\sigma$, $\omega$, $\rho$, and $\pi$; we use couplings as derived within the model and treat contact terms differently. 

This paper is organised as follows.  The Hartree-Fock QMC model is presented
in Sec.~\ref{sec:hadronic} and then used in Sec.~\ref{sec:gbem} to study hadronic matter in generalised beta-equilibrium. The proper time regularised NJL 
model is used to study three flavour quark matter in Sec.~\ref{sec:quark}. In Sec.~\ref{sec:crossover}, using
these two models a hybrid equation of state is obtained assuming a faux crossover
construction. In Sec.~\ref{sec:summary}, we summarise the results obtained and draw our conclusions.

\section{Hadronic model with quark degrees of freedom: The QMC model}
\label{sec:hadronic}

In this section we present the formalism used for hadronic matter in beta-equilibrium with leptons using the Hartree-Fock QMC model~\cite{PhysRevC.89.065801}. We will briefly review the derivation of the equations given in Refs.~\cite{PhysRevC.89.065801,MyThesis} and the approximations used therein.
All our parameters are fixed at saturation density in Symmetric (N=Z) Nuclear Matter. 
We then extrapolate the model
to investigate high density matter in generalised beta-equilibrium (GBEM),
which is relevant to neutron stars.

\subsection{The Lagrangian and Hamiltonian density}
\label{subsec:LagDen}

In our calculations we consider only the 
spin-$1/2$ octet baryons.
These baryons interact via the exchange of mesons which
couple directly to the quarks. The exchanged mesons included are the scalar-isoscalar ($\sigma$),
vector-isoscalar ($\omega$), vector-isovector ($\rho$),
 and pseudo-vector-isovector ($\pi$) bosons. These mesons only couple with the light quarks
 by the phenomenological OZI rule. We include the 
full vertex structure  for the vector mesons,  
that is, we include both the Dirac and Pauli terms.

The QMC Lagrangian density used in this work is given by a combination
of baryon and meson components
\begin{equation}
\mathcal{L} = \sum_{B}\mathcal{L}_{B} 
+ \sum_m \mathcal{L}_{m}\ ,
\label{eq:wholeL}
\end{equation}
for the octet of baryons \mbox{$B \in
  \{N,\Lambda,\Sigma,\Xi\}$}  and selected
mesons \mbox{$m \in \{\sigma,\omega,\rho,\pi\}$}
with the individual Lagrangian
densities
\begin{eqnarray}
\mathcal{L}_{B} & = &\label{eq:lb}
\bar{\Psi}_{B} \bigg(
i\gamma_{\mu}\partial^{\mu} - M_{B} + g_{\sigma B}(\sigma)\sigma \nonumber \\
&&- g_{\omega B}\gamma^{\mu}\omega_{\mu} -\frac{f_{\omega B}}{2M_{\rm N}}\sigma^{\mu\nu}\partial_{\mu}\omega_{\nu} \nonumber \\
&&  - g_{\rho B}\gamma^{\mu}\boldsymbol{t}\cdot\boldsymbol{\rho}_{\mu} -\frac{f_{\rho B}}{2M_{\rm N}}\sigma^{\mu\nu}\boldsymbol{t}\cdot\partial_{\mu}\boldsymbol{\rho}_{\nu} \nonumber \\
&&-\frac{g_{\rm A}}{2f_{\pi}}\chi_{BB}\gamma^{\mu}\gamma^{5}\boldsymbol{\tau}\cdot\partial_{\mu}\boldsymbol{\pi}
 \bigg) \Psi_{B}\ ,
\end{eqnarray}
\begin{eqnarray}
\sum_{m}\mathcal{L}_{m} &=& \frac{1}{2}(\partial_{\mu}\sigma\partial^{\mu}\sigma 
- m_{\sigma}^{2}\sigma^{2}) 
- \frac{1}{4}\Omega_{\mu\nu}\Omega^{\mu\nu} 
+ \frac{1}{2}m_{\omega}^{2}\omega_{\mu}\omega^{\mu} \nonumber \\
&& 
- \frac{1}{4}\boldsymbol{R}_{\mu\nu}\cdot\boldsymbol{R}^{\mu\nu} 
+ \frac{1}{2}m_{\rho}^{2}\boldsymbol{\rho}_{\mu}\cdot\boldsymbol{\rho}^{\mu} \nonumber\\
&&+ \frac{1}{2}(\partial_{\mu}\boldsymbol{\pi}\cdot\partial^{\mu}\boldsymbol{\pi} 
- m_{\pi}^{2}\boldsymbol{\pi}\cdot \boldsymbol{\pi})\ ,
\label{eq:lm}
\end{eqnarray}
for which the vector meson field strength tensors are 
$\Omega_{\mu\nu}=\partial_{\mu}\omega_{\nu}-\partial_{\nu}\omega_{\mu}$
and
$\boldsymbol{R}_{\mu\nu}=\partial_{\mu}\boldsymbol{\rho}_{\nu}-\partial_{\nu}\boldsymbol{\rho}_{\mu}$.
The $g_{iB}$, $f_{iB}$ denote meson-baryon coupling constants for the $i\in \left\lbrace \sigma , \omega , \rho\right\rbrace$ mesons.
The quantities in bold are vectors in isospin space with isospin matrices denoted by $\boldsymbol{t}$ and isospin Pauli matrices by $\boldsymbol{\tau}$. For nucleons and 
cascade particles $\boldsymbol{t} =\frac{1}{2}\boldsymbol{\tau}$.  
The pion-baryon interaction used here is assumed to be described by an
SU(3) invariant Lagrangian with the mixing parameter $\alpha = 2/5$~\cite{RikovskaStone:2006ta}
from which the hyperon-pion coupling constants can be given in terms of
the pion nucleon coupling,  $g^{\rm p.v.}_{\pi BB'} = g^{\rm p.v.}_{\pi NN} \chi_{BB'}=\frac{g_{A}}{2f_{\pi}}\chi_{BB'}$
\cite{RevModPhys.35.916,RikovskaStone:2006ta,Massot2008}, where $g_{A}$ and $f_{\pi}$ are the axial vector coupling and the pion decay 
constant, respectively.

From the Lagrangian given in Eq.~(\ref{eq:wholeL}) we obtain through the Euler-Lagrange equations a system of coupled non-linear partial differential equations for the quantum fields.
This is a difficult system of equations to solve and to make the problem tractable
 approximations are applied.  Static, no sea and mean field approximations are typically used and are implemented here. However, when we consider the NJL model we include 
the Dirac sea of negative energy states.

From the Hamiltonian density the EoS of nuclear matter can be calculated.
The Hamiltonian density in the static approximation is 
\begin{equation}
H = \int d^{3}r \left\lbrace  \mathcal{K}\  + \sum_{m\in\left\lbrace \sigma,\omega,\rho,\pi\right\rbrace } \mathcal{H}_{m} \right\rbrace \quad ,
\end{equation}
where we have decomposed it into its baryon and meson components as
\begin{eqnarray}
%%% K
\mathcal{K} & = &  \sum_{B} 
\bar{\Psi}_{\rm B} \left[  -i \vec{\gamma}\cdot\vec{\nabla} + M_{\rm B} 
%\sigma
- g_{\sigma B}(\sigma)\sigma \right] \Psi_{B}\quad , \\
%% H SIGMA
\mathcal{H}_{\sigma} & =& \frac{1}{2}\vec{\nabla}\sigma \cdot \vec{\nabla}\sigma +\frac{1}{2}m_{\sigma}^{2}\sigma^{2}  \quad ,\\
%% H OMEGA
\mathcal{H}_{\omega} & =& 
\sum_{B} 
\bar{\Psi}_{\rm B} \left[  
%omega vector
    g_{\omega B}\gamma_{\mu}\omega^{\mu}  
%omega tensor  
- \frac{f_{\omega B}}{2M_{\rm N}}\sigma_{\mu i}\partial^{i}\omega^{\mu}  
 \right] \Psi_{\rm B}  \nonumber\\
 & &   
 -\frac{1}{2}\left[ \vec{\nabla}\omega_{\mu}\cdot\vec{\nabla}\omega^{\mu} + (\vec{\nabla}\cdot\vec{\omega})^{2}+m_{\omega}^{2}\omega_{\mu}\omega^{\mu} \right] \, ,\\
 %% H RHO
\mathcal{H}_{\rho} & =& 
\sum_{B} 
\bar{\Psi}_{\rm B} \left[  
%rho vector 
   g_{\rho B}\gamma_{\mu}\boldsymbol{t}\cdot\boldsymbol{\rho}^{\ \mu} 
%rho tensor 
 - \frac{f_{\rho B}}{2M_{\rm N}}\sigma_{\mu i}
\boldsymbol{t}\cdot\partial^{i}\boldsymbol{\rho}^{\ \mu} \right] \Psi_{\rm B}  \nonumber\\
& & 
-\frac{1}{2}\left[ \vec{\nabla}\boldsymbol{\rho}_{\mu}\cdot\vec{\nabla}\boldsymbol{\rho}^{\mu} + (\vec{\nabla}\cdot\vec{\boldsymbol{\rho}})^{2}+m_{\rho}^{2}\boldsymbol{\rho}_{\mu}\cdot\boldsymbol{\rho}^{\mu} \right] \, ,\\
%%PION
\mathcal{H}_{\pi} & =& 
-\sum_{B} 
\bar{\Psi}_{\rm B} \left[  
%pion
\frac{g_{\rm A}}{2f_{\pi}}\chi_{BB} \gamma_{5}\boldsymbol{\tau}\cdot(\vec{\gamma}\cdot\vec{\nabla})\boldsymbol{\pi} 
\right] \Psi_{\rm B}  \nonumber\\
&& + \frac{1}{2}\vec{\nabla}\boldsymbol{\pi}\cdot\vec{\nabla}\boldsymbol{\pi} +\frac{1}{2}m_{\pi}^{2}\boldsymbol{\pi}\cdot\boldsymbol{\pi} \, . 
\end{eqnarray}

\subsection{Hartree-Fock approximation}
\label{subsec:HFA}

To perform the Hartree-Fock approximation 
we follow Refs.~\cite{Guichon:2006er,RikovskaStone:2006ta,Massot2008,HuToki:2010} by considering each meson field to be decomposed into two parts, a mean field part $\langle\phi \rangle$ and a fluctuation part $\delta\phi$, such that $\phi = \langle\phi\rangle +\delta\phi$ and solve the equations of motion order by order. The fluctuation terms are to be considered small with respect to the mean field contribution, the exception to this being the $\pi$ and $\rho$ meson fluctuations.
In this fashion, the $\sigma$ meson equation of motion is decomposed according to
\begin{eqnarray}
\left( -\vec{\nabla}^{2} \right. && + \left.   m_{\sigma}^{2}\right) (\bar{\sigma} +\delta\sigma) \label{eq:sig_eom_HF1}\\
&& =  -\frac{\partial \mathcal{K}}{\partial\sigma} 
 =  -\frac{\partial \mathcal{K}}{\partial\sigma}(\bar{\sigma}) -\delta\sigma\frac{\partial^{2} \mathcal{K}}{\partial\sigma^{2}}(\bar{\sigma}) - \ldots \, . \nonumber
\end{eqnarray}
The terms on the r.h.s of Eq.~(\ref{eq:sig_eom_HF1}) are expanded about their ground state expectation values (denoted by $\langle \ldots \rangle$). For the first term, we have   
\begin{eqnarray}
\frac{\partial \mathcal{K}}{\partial\bar{\sigma}} \equiv \frac{\partial \mathcal{K}}{\partial \sigma}(\bar{\sigma}) & = & \left\langle \frac{\partial \mathcal{K}}{\partial\bar{\sigma}} \right\rangle 
+ \delta \left[\frac{\partial \mathcal{K}}{\partial\bar{\sigma}}  \right]  \\
& = & \left\langle \frac{\partial \mathcal{K}}{\partial\bar{\sigma}} \right\rangle  + \left( \frac{\partial \mathcal{K}}{\partial\bar{\sigma}} -\left\langle \frac{\partial \mathcal{K}}{\partial\bar{\sigma}} \right\rangle \right)  
\end{eqnarray}
and similarly for the second and so on.
 We now proceed to solve the $\sigma$ meson equation of motion order by order. At the mean field or Hartree level we obtain
\begin{equation}
\left( -\vec{\nabla}^{2} +m_{\sigma}^{2}\right)\bar{\sigma} = -\left\langle  \frac{\partial \mathcal{K}}{\partial\bar{\sigma}} \right\rangle = \sum_{B} \left( -\frac{\partial M^{\ast}_{B}}{\partial\bar{\sigma}} \left\langle\bar{\Psi}_{B}\Psi_{B}\right\rangle\right)  
\end{equation}
and at the Fock level
\begin{eqnarray}
\left( -\vec{\nabla}^{2} +m_{\sigma}^{2}\right)\delta\sigma 
& = & -\left( \frac{\partial \mathcal{K}}{\partial\bar{\sigma}}  -\left\langle\frac{\partial \mathcal{K}}{\partial\bar{\sigma}} \right\rangle\right) \\ && -\delta\sigma \left[ \left\langle\frac{\partial^{2} \mathcal{K}}{\partial\bar{\sigma}^{2}}\right\rangle
 +\left( \frac{\partial^{2} \mathcal{K}}{\partial\bar{\sigma}^{2}} - \left\langle\frac{\partial^{2} \mathcal{K}}{\partial\bar{\sigma}^{2}} \right\rangle\right)  \right]  \nonumber \\
& = & - \frac{\partial \mathcal{K}}{\partial\bar{\sigma}}  +\left\langle\frac{\partial \mathcal{K}}{\partial\bar{\sigma}} \right\rangle  -\delta\sigma \frac{\partial^{2} \mathcal{K}}{\partial\bar{\sigma}^{2}}  \quad ,
\end{eqnarray}
where to this order ($\delta\sigma$)
\begin{equation}
\frac{\partial^{2} \mathcal{K}}{\partial\bar{\sigma}^{2}} \xrightarrow{\quad\quad}  \left\langle\frac{\partial^{2} \mathcal{K}}{\partial\bar{\sigma}^{2}}\right\rangle \quad .
\label{eq:toexpectd2Kds2}
\end{equation}
The fluctuation equation of motion becomes
\begin{equation}
\left( -\vec{\nabla}^{2} +m_{\sigma}^{2}\right)\delta\sigma =
- \frac{\partial \mathcal{K}}{\partial\bar{\sigma}}  +\left\langle\frac{\partial \mathcal{K}}{\partial\bar{\sigma}} \right\rangle  -\delta\sigma \left\langle\frac{\partial^{2} \mathcal{K}}{\partial\bar{\sigma}^{2}}\right\rangle \quad .
\label{eq:fl_eom_dKds}
\end{equation}
Eq.~(\ref{eq:fl_eom_dKds}) can be re-expressed in terms of an in-medium $\sigma$-meson mass and the fluctuation of the scalar baryon current as
\begin{equation}
\left( -\vec{\nabla}^{2} +m^{\ast\ 2}_{\sigma}\right)\delta\sigma 
 =  \sum_{B}  -\frac{\partial M^{\ast}_{B}}{\partial\bar{\sigma}} \left( \bar{\Psi}_{B}\Psi_{B} -\left\langle\bar{\Psi}_{B}\Psi_{B}\right\rangle\right) \quad ,
 \label{eq:fl_sig_eom_psi}
\end{equation}
where 
\begin{equation}
m^{\ast\ 2}_{\sigma} =  m_{\sigma}^{2} + \left\langle \frac{\partial^{2}\mathcal{K}}{\partial\bar{\sigma}^{2}}\right\rangle = m_{\sigma}^{2} + \sum_{B}\frac{\partial^{2}M^{\ast}_{B}}{\partial\bar{\sigma}^{2}} \left\langle \bar{\Psi}_{B}\Psi_{B}\right\rangle \quad .
\end{equation}

This in-medium $\sigma$ meson mass is only relevant to the fluctuating part and does not appear in the mean field portion of the $\sigma$ meson's equation of motion. This in-medium modification due to the baryons internal structure was included in Refs.~\cite{Guichon:2006er,RikovskaStone:2006ta,Massot2008}, but we will omit it here. We are neglecting this in-medium modification as we are approximating the Fock terms in the static approximation, omitting all other meson retardation effects and implementing a crude method of subtracting the contact terms that arise in the Fock terms. For these reasons it is reasonable to disregard it and use the free $\sigma$ meson mass in the Fock term, thereby treating it in the same manner as the other mesons.

The expectation value of the $\sigma$ field is given by
\begin{equation}
\bar{\sigma} = - \frac{1}{m_{\sigma}^{2}} \left\langle \frac{\partial \mathcal{K}}{\partial \bar{\sigma}} \right\rangle
= - \frac{1}{m_{\sigma}^{2}} \sum_{B} \frac{\partial M^{\ast}_{B}}{\partial \bar{\sigma}} \left\langle \bar{\Psi}_{B}\Psi_{B}\right\rangle \quad ,
\label{eq:sigma_bar}
\end{equation}
which is then determined numerically. Krein {\it et al.}~\cite{Krein:1998vc} also considered an additional correction involving the mean scalar field generated by the Fock terms. This can be done by considering the energy density as a functional and requiring it to be thermodynamically consistent, meaning that the total energy density, $\epsilon$, is minimised with respect to $\bar{\sigma}$.
This amounts to Eq.~(\ref{eq:sigma_bar}) plus an additional term because of the dependence of the Fock contribution to the energy density on $\bar{\sigma}$.

The fluctuation of the $\sigma$ field can now be written in terms of the $\sigma$ meson's Green function $\Delta_{\sigma}$ as
\begin{eqnarray}
\delta\sigma(\vec{r}) & = & \int d^{3}r'\ \Delta_{\sigma}(\vec{r} -\vec{r}\ ') \left( -\frac{\partial \mathcal{K}}{\partial\bar{\sigma}} +\left\langle\frac{\partial \mathcal{K}}{\partial\bar{\sigma}}\right\rangle \right) (\vec{r}\ ' ) \quad\quad \quad \\ 
& = & 
\label{eq:delta_sig_psi}
\sum_{B} \int d^{3}r' \frac{d^{3}q}{(2\pi)^{3}}\ e^{i\vec{q}\cdot (\vec{r}-\vec{r}\ ')}\Delta_{\sigma}(\vec{q}) \\
&& \quad\quad  \left[ - \frac{\partial M^{\ast}_{B}}{\partial\bar{\sigma}}\left(  \bar{\Psi}_{B}\Psi_{B}-\left\langle\bar{\Psi}_{B}\Psi_{B}\right\rangle\right) (\vec{r}\ ') \right]\, . \nonumber   
\end{eqnarray}

By considering the meson fields decomposed into a mean field part and a fluctuating part it can be seen in Eq.~(\ref{eq:fl_sig_eom_psi}) and (\ref{eq:delta_sig_psi}), that this is related to a similar decomposition of the baryon currents. We introduce the following notation:
\begin{eqnarray}
\bar{\Psi}_{B}\tilde{\Gamma}_{\alpha B}\Psi_{B} & =  &\left\langle \bar{\Psi}_{B}\tilde{\Gamma}_{\alpha B}\Psi_{B}\right\rangle  \\
&& \,+ \,  (\bar{\Psi}_{B}\tilde{\Gamma}_{\alpha B}\Psi_{B} - \left\langle \bar{\Psi}_{B}\tilde{\Gamma}_{\alpha B}\Psi_{B}\right\rangle ) \nonumber\\
& =  &\left\langle \bar{\Psi}_{B}\tilde{\Gamma}_{\alpha B}\Psi_{B}\right\rangle  
+ \delta(\bar{\Psi}_{B}\tilde{\Gamma}_{\alpha B}\Psi_{B} )  ,
\end{eqnarray}
where $\tilde{\Gamma}_{\alpha B}$ denotes one of the meson-baryon interactions, appearing in the Lagrangian (Eq.~(\ref{eq:lb})), written in terms of the appropriate Lorentz and isospin structures.

The analogous expressions for the remaining mesons follow in same manner. Here we summarise the equations of motion for all mesons decomposed into mean field and  fluctuation components.
For the meson mean fields we have
\begin{subequations}
\label{allequations}
\begin{eqnarray}
\bar{\sigma} 
& = & - \frac{1}{m_{\sigma}^{2}} \sum_{B} \frac{\partial M^{\ast}_{B}}{\partial \bar{\sigma}} \left\langle \bar{\Psi}_{B}\Psi_{B}\right\rangle \ , \\
\bar{\omega} & = &\frac{1}{m^{2}_{\omega}} \sum_{B}   g_{\omega B}\left\langle \Psi^{\dagger}_{\rm B} \Psi_{\rm B}\right\rangle \  ,\\
\bar{\rho} & = &\frac{1}{m^{2}_{\rho}} \sum_{B}   g_{\rho B}\left\langle \Psi^{\dagger}_{\rm B} t_{3B}\Psi_{\rm B}\right\rangle \  , 
\end{eqnarray}
\end{subequations}
and $\bar{\pi}  =  0$.
Each meson field fluctuation can be condensed to 
\begin{equation}
\delta\phi (\vec{r}\, ) = \sum_{B}\int d^{3}r' \ \Delta_{\phi}(\vec{r} - \vec{r}\, ' ) \delta (\bar{\Psi}_{\rm B}\tilde{\Gamma}_{\phi B}\Psi_{\rm B} )(\vec{r}\, ') \quad ,
\label{eq:meson_fluctuation}
\end{equation}
where $\Delta_{\phi}$ is the Yukawa propagator for the meson $\phi\in \left\lbrace  \sigma,\omega,\rho,\pi \right\rbrace$.

The decomposition of the meson fields also occurs in the Hamiltonian.   
For example, the baryon kinetic and  the $\sigma$ meson terms amount to
\begin{eqnarray}
&&\int d^{3}r \   \left[  \mathcal{K} + \mathcal{H}_{\sigma} \right]   \\
& = & \int  d^{3}r  \left[ \mathcal{K}(\bar{\sigma}) + \frac{1}{2}\delta\sigma \left( \frac{\partial \mathcal{K}}{\partial\bar{\sigma}}  -  \left\langle \frac{\partial\mathcal{K}}{\partial \bar{\sigma}}\right\rangle\right)
+ \frac{1}{2}m_{\sigma}^{2}\bar{\sigma}^{2} \right]\nonumber\\   
& =& \int  d^{3}r  \left[ 
\sum_{B}\bar{\Psi}_{\rm B} \left[  -i \vec{\gamma}\cdot\vec{\nabla} + M_{\rm B} 
%\sigma
- g_{\sigma B}(\bar{\sigma})\bar{\sigma} \right] \Psi_{B}  \nonumber \right.\\
&& \left. + \frac{1}{2} \sum_{B}\frac{\partial M^{\ast}_{\rm B}}{\partial\bar{\sigma}} \delta\sigma(\vec{r}) \delta \left(\bar{\Psi}_{\rm B}\Psi_{\rm B}\right)(\vec{r})  + \frac{1}{2}m_{\sigma}^{2}\bar{\sigma}^{2} \right] \quad . 
\end{eqnarray}
The first term in the last line is the $\sigma$ meson's Fock term contribution to the energy.
To proceed further, we must first explain how the in-medium Dirac equation for the baryons is solved in the Hartree-Fock approximation. This is presented in Sec.~\ref{subsec:inmed_DEq}. In the process of doing this, we Fourier transform to momentum space, where the energy density of nuclear matter is more easily evaluated.

\subsection{The in-medium Dirac equation}
\label{subsec:inmed_DEq}

The in-medium Dirac equation, for a baryon $i$ in nuclear matter, can be written as
\begin{equation}
\left( \slashed{p} - M_{i} - \Sigma_{i}(p) \right) u_{i}(p,s) = 0 \quad ,
\end{equation}
where $\Sigma_{i}(p)$ is the self-energy of the baryon. From parity conservation and translational, rotational and time reversal invariance, the self-energy can be decomposed into three scalar functions in the nuclear matter rest frame~\cite{Serot:1984ey},
\begin{equation}
\Sigma_{i}(p) = \Sigma_{i}^{\rm s}(p) + \gamma^{0}\Sigma_{i}^{\rm 0}(p) + \vec{\gamma}\cdot\vec{p}\, \Sigma_{i}^{\rm v}(p) \quad .
\label{eq:selfenergy}
\end{equation}
The functions $\Sigma_{i}^{\rm s}(k)$, $\Sigma_{i}^{\rm 0}(k)$ and $\Sigma_{i}^{\rm v}(k)$ are the scalar, temporal vector and spatial vector components of the self-energy, respectively.

If we introduce the following effective quantities
\begin{eqnarray}
M^{\ast}_{i} (p) & = &  M_{i} + \Sigma_{i}^{\rm s}(p) \quad , \\
E_{i}^{\ast}(p) & = &  E_{i}(p) + \Sigma_{i}^{\rm 0}(p) = \sqrt{\vec{p}^{\, \ast\, 2} + M^{\ast\, 2}_{i}}  \quad ,\\
\vec{p}^{\, \ast} &= & \vec{p}\, ( 1 + \Sigma_{i}^{\rm v}(p) )\quad ,
\end{eqnarray}
the Dirac equation can be written in a form 
which is formally equivalent to the Dirac equation in vacuum. Therefore, as in vacuum, the positive energy solution to the in-medium Dirac equation is 
\begin{equation}
u_{i}(p,s) =\sqrt{\frac{M^{\ast}_{\rm i} + E^{\ast}_{\rm i}}{2E^{\ast}_{i}}} 
\left( \begin{array}{c}
1 \\
\frac{\vec{\sigma}\cdot\vec{p}^{\, \ast}}{M^{\ast}_{i} + E^{\ast}_{i}}
\end{array}\right) 
\chi_{s} \quad ,
\end{equation}
where $\chi_{s}$ are Pauli spinors and we have used the normalisation convention $u_{i}^{\dagger}(p,s)u_{i}(p,s) = 1 $ for the spinor~\cite{Serot:1984ey}.

From fully self-consistent calculations performed using QHD~\cite{Serot:1984ey,Bouyssy:1987} and QMC~\cite{Krein:1998vc}  models, it is known that the scalar and temporal vector self-energy components are approximately momentum independent and the spatial vector component is very small. Therefore, we proceed by carrying out the self-consistency approximately, as in Ref.~\cite{MyThesis,PhysRevC.89.065801}, where we ignore these small contributions. The self-energy then has a form identical to the usual mean-field (Hartree) result and the small Fock corrections can be included by requiring thermodynamic consistency.

The no sea approximation is used, i.e., the negative energy states of the baryons are ignored. Therefore, the in-medium propagators for baryons propagating on-shell in the nuclear matter rest frame are given entirely by the Dirac portion of the baryon propagator~\cite{Serot:1984ey}
\begin{equation}
G_{\alpha\beta}^{i}(p) = \frac{i\pi}{E^{\ast}_{i}(p)} (\slashed{p}^{\ast} +M^{\ast}_{i})_{\alpha\beta}\delta(p^{0} - E(p))\Theta(p_{F,i}-\vert\vec{p}\vert ) \quad ,
\label{eq:inmed_baryon_prop}
\end{equation}
where $p_{\rm F,i}$ is the Fermi momentum of baryon $i$.

With the above approximations and definitions, the Fock contribution can be evaluated.  For the $\sigma$ meson Fock contribution to the energy density, we obtain
\begin{eqnarray}
\epsilon^{\rm F}_{\sigma} & = & \frac{1}{2}\sum_{i} \left( \frac{\partial M^{\ast}_{i}}{\partial\bar{\sigma}}\right)^{2}
\int\limits_{\vert\vec{p}\vert \leq p_{F,i}} \frac{d^{3}p}{(2\pi)^{3}} \int\limits_{\vert\vec{p}\, '\vert \leq p_{F,i}'} \frac{d^{3}p'}{(2\pi)^{3}} \nonumber\\
 && \quad \Delta_{\sigma}(\vec{q}\, )
 \frac{{\rm Tr}\left[ \left( \slashed{p}^{\, \ast} + M^{\ast}_{i}\right) \left( \slashed{p}^{\, \prime\ast} + M^{\ast}_{i}\right)\right] }{4 E^{\ast}_{i}(p)E^{\ast}_{i}(p')} \, .
 \label{eq:sigmaEFnoFF}
\end{eqnarray}
As can be seen in Eq.~(\ref{eq:sigmaEFnoFF}), there is an additional scalar dependence in this Fock term, which appears after explicit evaluation. A correction to the mean scalar field can easily be included numerically. This is a small contribution, as illustrated by the scenarios labelled ``Fock $\delta\sigma$" in Ref.~\cite{MyThesis,PhysRevC.89.065801}, and so we neglect it here. The Fock contributions of the other mesons are straightforwardly obtained in the same manner. They too have an additional dependence on the scalar mean field.

\subsection{Hartree-Fock equation of state}
\label{subsec:HFEoS}
Here, to describe the baryon-meson interaction, we also introduce form factors, because of the extended nature of the baryons, by 
\begin{equation}
g_{\alpha B} \xrightarrow{\quad\quad}g_{\alpha B} F^{\alpha}(k^{2}) \quad .
\end{equation}
The $\sigma$, $\omega$, $\rho$ and $\pi$ form factors are all taken to have
the dipole form $F(k^{2}) \simeq F(\vec{k}^{2})$ with the same
cut-off $\Lambda$. We explored values of the cut-off mass in the range 0.9 -- 1.3~GeV in Ref.~\cite{PhysRevC.89.065801}. Clearly, these form factors are only of concern for
the Fock terms as these allow for a finite momentum transfer, 
whereas Hartree contributions do not.

Within the QMC model, the hadronic energy density $\epsilon_{hadronic}$ is the sum of the of the baryonic
energy density in nuclear matter which is 
\begin{equation}
\epsilon_{B} = \frac{2}{(2\pi)^{3}}\sum_{B} 
\int\limits_{\vert \vec{p}\vert \leq p_{F}}d^{3}p\ 
\sqrt{\vec{p}^{\, 2}+M^{\ast\, 2}_{B}} \, , \  
\end{equation}
and the mesonic energy density $\epsilon_{\sigma \omega \rho\pi}$. This can be
divided into two parts, the Hartree $\epsilon_{H}=\epsilon_{B} 
+\epsilon_{\sigma\omega\rho}^{H} $ and the Fock 
$\epsilon_{F}=\epsilon^{F}_{\sigma\omega\rho} + \epsilon_{\pi}$ contribution.
The total mesonic energy density is given by $\epsilon_{\sigma \omega \rho\pi}
= \epsilon^{H}_{\sigma\omega\rho} + \epsilon_{F}$, where the Hartree and Fock
components of the mesonic energy density are given respectively by
\begin{equation}
\epsilon^{H}_{\sigma\omega\rho} = \sum_{\alpha\in\left\lbrace \sigma,\omega,\rho\right\rbrace }\frac{1}{2}m^{2}_{\alpha}
\bar{\alpha}^{2} 
\end{equation}
where $\bar{\alpha}$ refers to the mean value 
of meson field $\alpha$ and 
\begin{eqnarray}
\epsilon_{F} & = & \frac{1}{(2\pi)^{6}}\sum_{m\in\left\lbrace \sigma,\omega,\rho,\pi\right\rbrace }\sum_{BB'} 
C^{m}_{BB'}  \nonumber\\
&& \hspace*{10mm} \int\limits_{\substack{\vert \vec{p}\vert \leq p_{F}}}
\int\limits_{\substack{\vert \vec{p}\, '\vert \leq p_{F'}}} d^{3}p d^{3}p' \ 
\mathbf{\Xi}_{BB'}^{\alpha} \ ,
\end{eqnarray}
where $C^{\sigma}_{BB'}=C^{\omega}_{BB'}=\delta_{BB'}$.  
$C^{\rho}_{BB'}$ and $C^{\pi}_{BB'}$, which arise 
from symmetry considerations, 
are given in Ref.~\cite{RikovskaStone:2006ta} and $\Xi_{BB'}^{\alpha}$, is explained below.  
The additional contribution from the Fock terms to the scalar self-consistency significantly increases
computation time and was shown to make only a small change in
our results in Ref.~\cite{PhysRevC.89.065801}. For this reason we neglect its correction to the $\sigma$ mean field.
The vector meson mean fields are given by 
\begin{equation}
\bar{\omega} =\sum_{i} \frac{g_{\omega i}}{m^{2}_{\omega}} \rho_{i}^{\rm v} \quad \textrm{ and } \quad 
\bar{\rho} =\sum_{i} \frac{g_{\rho i}}{m^{2}_{\rho}}t_{3i} \rho_{i}^{\rm v} \quad ,
\end{equation}
where $\rho^{\rm v}_{i}$ is the number density for baryon $i$.

For $\epsilon^{F}$, the integrand has the form 
\begin{equation}
\mathbf{\Xi}^{m}_{BB'}  = \frac{1}{2}\sum_{s,s'} \vert
\bar{u}_{B'}(p',s')\Gamma_{mB} u_{B}(p,s)\vert^{2} 
\Delta_{m}(\vec{k})\ ,
\end{equation}
where $\Delta_{m}(\vec{k})$ is the Yukawa propagator for 
meson $m$ with momentum $\vec{k} =\vec{p} - \vec{p'}$  
and $u_{B}$ are the baryon spinors.
In the above integrands we expand to isolate the momentum independent pieces and simply
subtract contact terms as in Ref.~\cite{PhysRevC.89.065801}. We emphasise here the importance of subtraction
of the momentum independent piece, which when transformed to configuration space corresponds to a delta
function.The removal of the contact terms is a common procedure because
they represent very short range correlations between the
baryons. To keep them in this model would not be consistent as it treats the baryons as clusters of quarks and not as point-like
objects. Moreover, it is also required because of the suppression of the relative wave function at short distance originating from the repulsive hard core.

\subsection{QMC model parameters}
\label{subsec:Model_Params}

{\squeezetable
 \begin{table}%[H] add [H] placement to break table across pages  
 \begin{ruledtabular}
 \begin{tabular}{lcccccccc}
 \multirow{2}{*}{Model/} & \multirow{2}{*}{\ $g_{\sigma N}$\ } &\multirow{2}{*}{\ $g_{\omega N}$\ }&\multirow{2}{*}{\ $g_{\rho}$\ }& $K_{0}$&$L_{0}$ & $M_{\rm max}$& $R$
  & $\rho^{\rm max}_{c}$\\
 Scenario&& & & [MeV] &[MeV]  & [$M_\odot$] & [km]& [$\rho_0$] \\[2mm]
 \hline\\
{\footnotesize Standard }  & 8.97 & 9.38 & 4.96 &273 & 84  & 1.80 & 11.80 & 5.88  \\   
{\footnotesize $\Lambda = 1.3$ }   & 9.31 & 10.67  & 5.40  &289 & 88  &  1.95& 12.10 & 5.52  \\   \end{tabular}
 \end{ruledtabular}
 \caption{ Couplings, nuclear matter properties, and neutron star properties determined for our standard 
case (for which $\Lambda=0.9$~GeV, and 
$R^{\rm free}_{N}=1.0$~fm) and the scenario where the cut-off is increased to $\Lambda = 1.3$~GeV. 
The symmetric nuclear matter quantities evaluated at saturation, $K_{0}$ and $L_{0}$, 
are the incompressibility and slope of the symmetry energy, respectively. 
Tabulated neutron star quantities are the stellar radius, maximum stellar mass and 
corresponding central density (units $\rho_0=0.16$~fm$^{-3}$).
\protect\label{table:couplings}}  
 \end{table}
}

The model dependence on all parameters was discussed in detail in Ref.~\cite{PhysRevC.89.065801}.
There are   
just the three main adjustable coupling constants, which control 
the coupling of the mesons to the two lightest quarks, $g^{q}_{\sigma}$,
$g^{q}_{\omega}$, and $g^{q}_{\rho}$ for $q=u,d$ 
($g^{s}_{\alpha}=0$ for all mesons $\alpha$). 
 In addition, one has the
meson masses, the value of the cut-off parameter 
$\Lambda$ appearing in the dipole form factors needed to evaluate 
the Fock terms and finally 
the bag radius of the free nucleon.
The $\sigma$, $\omega$, and $\rho$ couplings to the quarks are
constrained to reproduce the standard empirical properties of 
symmetric ($N$=$Z$) nuclear matter; the saturation 
density $\rho_{\rm 0}$ = 0.16 fm$^{\rm -3}$, the 
binding energy per nucleon at saturation 
of $\mathcal{E}(\rho = \rho_{0}) = -15.865$~MeV 
as well as 
the asymmetry energy coefficient \mbox{$a_{\rm asym} \equiv
  S_{0} \equiv S(\rho_{0}) = 32.5$~MeV}~\cite{RikovskaStone:2006ta}.

The $\omega$, $\rho$ and $\pi$ meson masses are set to 
their experimental values.  
Whereas, the value of the $\sigma$ mesons mass is
taken to be $700$~MeV~\cite{Guichon:1995ue,Guichon:2006er,PhysRevC.89.065801}.
The form factor cut-off mass, $\Lambda$, controls the
strength of the Fock terms. 
In Ref.~\cite{PhysRevC.89.065801} the preferred value was ~0.9 GeV. Here we consider only two values, the preferred
0.9 GeV and 1.3 GeV. The latter was chosen as it produces an overly stiff EoS.
All the other coupling constants in the expression for the total energy 
density are \textit{calculated} within the QMC model or determined from 
symmetry considerations without further need for adjustable parameters. 
In particular, the tensor couplings are determined from experimental magnetic moments.

\section{Generalised beta-equilibrium matter and neutron stars}
\label{sec:gbem}

As the density of hadronic matter increases beyond saturation density nuclei
 dissolve to form an interacting system of nucleons and leptons. If this system survives 
 longer than the time scale of weak interactions, 
$\tau \approx$ 10$^{\rm -10}$~s, it is able to reach equilibrium with respect to 
beta decay $n \rightarrow p + e^{\rm -} + \tilde{\nu}$ 
 and its inverse .
 When the total baryonic density reaches about 2 -- 3 $\rho_{\rm 0}$ and 
because baryons obey the Pauli exclusion principle, it becomes energetically more favourable to 
create a slow and more massive hyperon, rather than another energetic nucleon. A generalised beta equilibrium  develops with respect to all reactions involving either
weak or strong interactions, that leads to the lowest energy state.  
Only two quantities are conserved 
in GBEM\textemdash the total charge (zero in stars) and total baryon number. 
Strangeness is conserved only on the time scale of strong interaction, 
$\tau \approx$ 10$^{\rm -24}$~s, and lepton number is conserved only on 
the time-scale of tens of seconds, because of the diffusion of 
neutrinos out of the star~\cite{Glendenning:1997wn}.

 We supplement the
QMC model developed so far with non-interacting leptons and proceed to study matter in generalised beta-equilibrium (allowing for hyperons as well as nucleons).
Only electrons and muons are considered, as  tau leptons are too massive to be found in neutron stars. As we are considering old neutron stars, neutrinos are assumed to have radiated out of the star, so they can also be neglected. For the lepton masses we use their experimental values~\cite{Beringer:1900zz}.
The corresponding lepton energy density and number density are given by the usual formulas for a
degenerate Fermi gas~\cite{Glendenning:1997wn}.
The total energy density of the GBEM is then given by the sum of the hadron and lepton energy densities, $\epsilon_{\rm total} =
\epsilon_{\rm hadronic} + \epsilon_\ell$. Similarly the total pressure is the sum of the hadron and lepton pressures and can be calculated as
\begin{equation}
\label{eq:deftotpressure}
P_{\rm total} = \rho^{2}\frac{\partial}{\partial \rho}
\left(\frac{\epsilon_{\rm total}}{\rho}\right) 
= \sum_{i,\ell} \mu_{i}\rho_{i}-\epsilon_{\rm total} \quad  .
\end{equation} 

To describe GBEM, we need to determine the lowest energy state under the two constraints of baryon
number conservation and charge neutrality. For this 
we use the standard method of Lagrange multipliers. The equilibrium configuration of the system is then
determined variationally.

We solve the following system of equations: 
\begin{eqnarray}
0 &=& \mu_{i} + B_{i}\lambda + \nu Q_i\ , \label{eq:SysEq1}\\
0 &=& \mu_{\ell}-\nu\ ,  \label{eq:SysEq2}\\
0 &=& \sum_{i} B_{i} \rho^{\rm v}_{i}-\rho\ , \label{eq:SysEq3}\\
0 &=& \sum_{i} B_{i} \rho_{i}^{\rm v}Q_{i} 
+ \sum_{\ell} \rho_{\ell}Q_{\ell}\  ,  \label{eq:SysEq4}\ 
\end{eqnarray}
where the baryon number, $B_{i}$, is 0 or 1 and the charge number, $Q_{i}$, is 0 or $\pm$1.
This system of equations is solved to obtain the number densities
for each particle 
(\mbox{$i\in\{n,p,\Lambda,\Sigma^-,\Sigma^{0},\Sigma^+,\Xi^{-},\Xi^0\}$}
and \mbox{$\ell\in\{e^{-},\mu^{-}\}$}), $\rho_{i}^{\rm v}$, as well as 
the Lagrange multipliers ($\lambda$, $\nu$). 

At Hartree--Fock level, the following formulas to
numerically evaluate the chemical potentials, must be used to ensure we
encapsulate the Fock contribution to the energy densities correctly
\begin{equation}
\mu_{i} = \frac{\partial \epsilon_{\rm total}}{\partial \rho^{\rm v}_{i}}\ , 
\quad \mu_{\ell} = \frac{\partial\epsilon_{\ell}}{\partial \rho_{\ell}} 
= \sqrt{k^{\ell \, 2}_{F}+m^{2}_{\ell}} \, .
\label{eq:chempots}
\end{equation} 

The EoS of GBEM is not valid in the 
outer regions (crust) of the star, where nuclei and nuclear processes become dominant. Following 
the customary procedure, we introduce 
a smooth transition between our EoS in  GBEM  
and the standard low density EoS of Baym, Pethick and 
Sutherland (BPS)~\cite{Baym:1971pw} at 
low density.
In order to calculate neutron star properties, such as the total
gravitational mass, $M(R)$, within the stellar radius $R$, 
we solve the TOV equations~\cite{PhysRev.55.374,PhysRev.55.364,Tolman01031934} for hydrostatic 
equilibrium of spherically symmetric (non-rotating) matter.

\subsection{Hadronic matter numerical results}
\label{subsec:Hadronic_NumRes}

%%%% HADRONIC PLOTS
\begin{figure}
\includegraphics[width=0.45\textwidth]{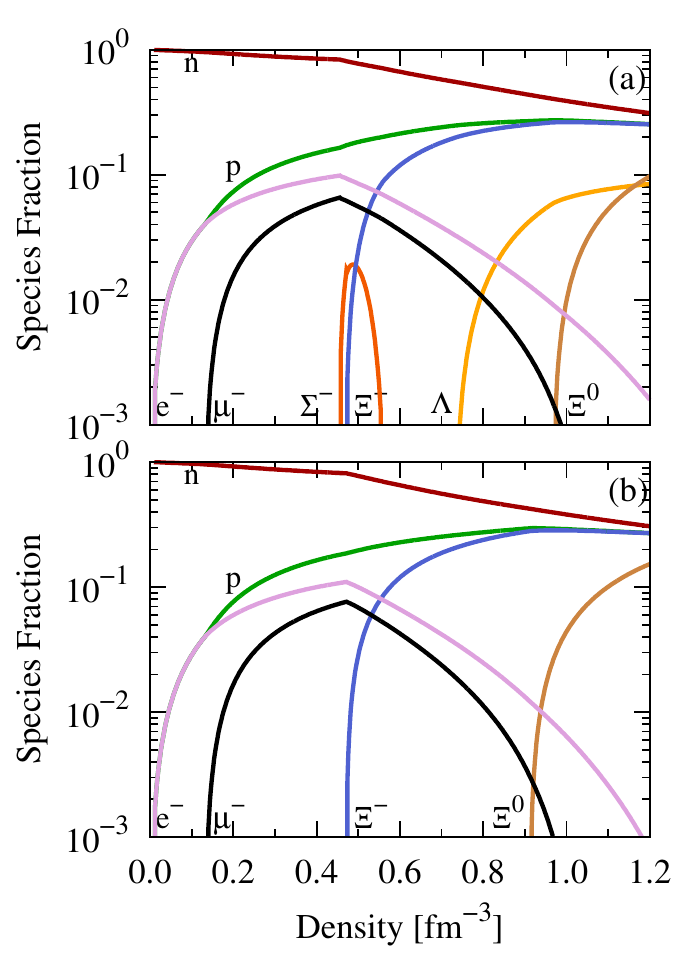} 
\caption{Species fractions as a function of density for hadronic matter in generalised beta-equilibrium for (a) Standard or baseline scenario and (b) $\Lambda =1.3$~GeV.  }
\label{fig:SpFr_Hadronic}
\end{figure}

\begin{figure}
\includegraphics[width=0.5\textwidth]{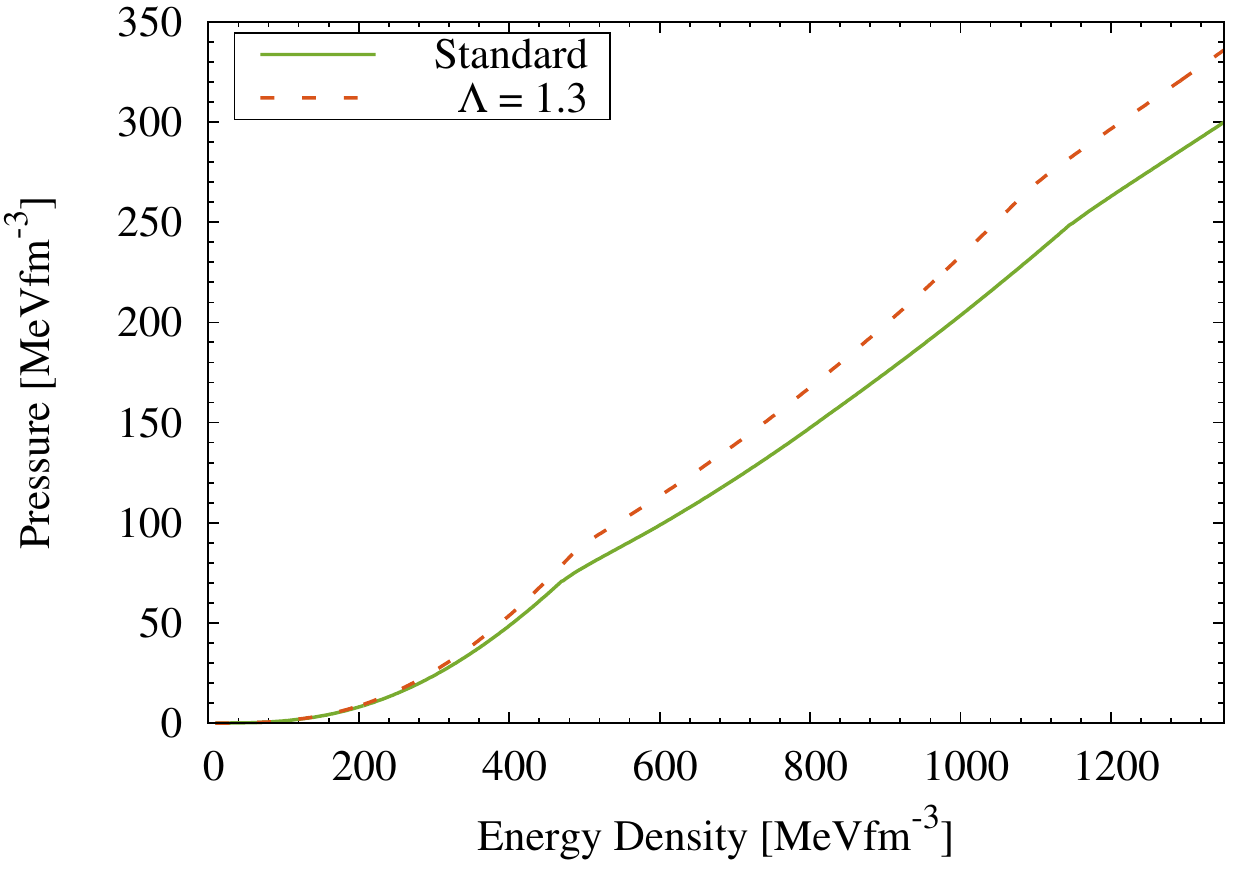} 
\caption{Pressure as a function of energy density for hadronic matter in generalised beta-equilibrium.}
\label{fig:Hadronic_PVsEnDen}
\end{figure}

Quark-meson coupling model numerical results were discussed in 
detail in Ref.~\cite{PhysRevC.89.065801}. Here, we simply present the numerical results
for two model variations, $\Lambda=0.9$~GeV and the overly stiff $\Lambda=1.3$~GeV.
Figure~\ref{fig:SpFr_Hadronic} shows the species fraction for hadronic matter in generalised beta-equilibrium. 
The corresponding EoS are depicted in Fig.~\ref{fig:Hadronic_PVsEnDen} .

\section{Quark model: Proper time regularised NJL model}
\label{sec:quark}

In this section, we introduce the proper time regularised three flavour Nambu\textendash Jona-Lasinio (NJL) model
and use it to study quark matter.

We use an NJL Lagrangian which is inspired by one gluon exchange and proceed to calculate the corresponding effective potential in the mean field approximation. 
One gluon exchange can be approximated by constructing a current-current interaction using the conserved colour current
$J^{\mu}_{a} = \bar{\psi}\gamma^{\mu}t_{a}\psi$. 
Performing Fierz transformations on this interaction Lagrangian in the exchanged $q\bar{q}$-channel and 
adding it to the original, one obtains an interaction Lagrangian consisting of colour singlet and octet terms. As we
are only interested  in colour neutral matter, we ignore the colour octet terms. 
The Lagrangian density we investigate is then:
\begin{eqnarray}
\label{eq:NJLLag}
\mathcal{L}_{\rm NJL} &=& \bar{\psi} (i\slashed{\partial}-\hat{m}_{0})\psi \nonumber\\
&& +\ G_{\rm S}\sum_{a=0}^{N_{\rm F}^{2}-1} \left[ \left( \bar{\psi}\lambda_{a}\psi\right)^{2}+ \left(\bar{\psi}i\gamma_{5}\lambda_{a}\psi \right)^{2} \right] \nonumber \\
& &   -\ G_{\rm V}\sum_{a=0}^{N_{\rm F}^{2}-1} \left[ \left( \bar{\psi}\gamma_{\mu}\lambda_{a}\psi\right)^{2}+ \left(\bar{\psi}\gamma_{\mu}\gamma_{5}\lambda_{a}\psi \right)^{2} \right] , \quad\quad
\end{eqnarray}
where $\hat{m}_{0}=$ diag$(m_{u},m_{d},m_{s})$. We used $t_{a}$ to denote the $SU(3)$ Gell-Mann matrices in colour space, whereas in Eq.~(\ref{eq:NJLLag}) we use $\lambda_{a}$ to label them in flavour space. 
Through the application of the Fierz transformations, one finds that the vector coupling is simply related to the scalar coupling, i.e., $G_{\rm V}=\frac{G_{\rm S}}{2}$. However, in practice it can be constrained by some physical quantity such as a vector meson mass. We will treat the vector coupling as a free parameter, varying it from zero up to equality with the scalar coupling, in order to understand its effect on the quark equation of state.

 The dynamically generated constituent quark masses in this NJL model in the MFA are then given by
$M_{i} = m_{i} - 4 G_{\rm S}\langle\bar{\psi}_{i}\psi_{i}\rangle$, 
where $m_{i} $ is the current quark mass of flavour $i\in\left\lbrace u,d,s \right\rbrace$. In vacuum, the explicit proper time regularised expression is 
\begin{equation}
\label{eq:ConstitQMassReg}
M_{i} = m_{i} + \frac{3G_{\rm S}M_{i}}{\pi^{2}}\int^{\infty}_{\frac{1}{\Lambda_{UV}^{2}}}d\tau \frac{1}{\tau^{2}}e^{-\tau M_{i}^{2}}\quad .
\end{equation}
At finite density Eq.~(\ref{eq:ConstitQMassReg}) gains an additional correction.

\subsection{NJL model parameters}
\label{subsec:njlparams}

{\squeezetable
 \begin{table*}%[H] add [H] placement to break table across pages
 \begin{ruledtabular}
 \begin{tabular}{cccccccccc}
Model  & $m_{\ell}$ & $m_{s}$  & $M_{\ell}$ & $M_{s}$ & $\Lambda_{\rm UV}$ & $G_{\rm S}$ & $G_{\rm D}$ & $\vert\langle \bar{\psi}_{\ell}\psi_{\ell}\rangle_{0}\vert^{1/3}$ & $\vert\langle \bar{\psi}_{s}\psi_{s}\rangle_{0}\vert^{1/3}$ \\
 & [MeV] & [MeV] & [MeV]& [MeV] & [MeV] & [GeV$^{-2}$] & [GeV$^{-5}$] & [MeV] & [MeV] \\
 \hline\\%\midrule
% PTR   &  \\ %Proper time reg.
{\footnotesize PS1}   & 17.08 & 279.81 & $400^{\ast}$ & $563^{\ast}$&  636.67& 19.76 & -& 169.20& 153.01 \\
 %&  \\
  {\footnotesize PS2}  &$5.5^{\ast}$ & $135.7^{\ast}$ &201.07& 440.41 & 1078.9 & 3.17 & - & 249.0 & 288.67\\
%\hline
%TMR   & &&&&&&&& \\ %Three mom. reg.
 {\footnotesize HK}~\cite{Masuda:2012ed,Hatsuda:1994pi}   &5.5 & 135.69 & 334.59 & 527.28 & 631.38 & 4.60 & 92.57 &246.72&266.94\\
 \end{tabular}
 \end{ruledtabular}
 \caption{We fitted all parameter sets using low energy hadron phenomenology. The proper time regularised (PTR) parameter sets used $f_{\pi}=93$~MeV and  $m_{\pi}=140$~MeV.  An asterisk ($\ast$) marks variables used in the fitting procedure. The three momentum regularised (TMR) parameter set HK~\cite{Masuda:2012ed,Hatsuda:1994pi} used $f_{\pi}=93$~MeV,  $m_{\pi}=138$~MeV, $m_{K} = 495.7$~MeV,  $m_{\eta\prime}=957.5$~MeV and $m_{\ell} = 5.5$~MeV. For the HK model, we have the additional coupling $G_{\rm D}$ which is the coupling strength of the determinant term.}
\label{table:NJLPSets}
 \end{table*}
}

The NJL model has essentially five model parameters, the UV cut-off $\Lambda_{\rm UV}$; the scalar coupling $G_{\rm S}$; the light and strange constituent quark masses $M_{\ell, s}$ or equivalently their current quark masses $m_{\ell,s}$ (there is a one-to-one relationship between constituent and current quark masses by Eq.~(\ref{eq:ConstitQMassReg})); and finally the vector coupling, $G_{\rm V}$, which is treated as a free parameter.
As is common practice, we use pion phenomenology to fit the NJL model parameters.
We constrain our model parameters in two ways. In parameter set PS1, we take as input the constituent quark masses $M_{\ell}=400$~MeV and $M_{s}=563$~MeV, the pion's mass $m_{\pi}=140$~MeV and its decay constant $f_{\pi}=93$~MeV. By requiring these values for $M_{\ell}$ and $f_{\pi}$, the UV cut-off, $\Lambda_{\rm UV}$, is constrained to be $636.67$~MeV. Then using the correspondence of the pole in the quark-anti-quark t-matrix in the pseudoscalar channel to the physical pion mass $m_{\pi}$, along with the values $M_{\ell}$ and $\Lambda_{\rm UV}$ the scalar coupling $G_{\rm S}$ is found to be $19.76$~GeV$^{-2}$. Finally, using $M_{\ell, s}$, $\Lambda_{\rm UV}$ and $G_{\rm S}$ the current quark masses can be calculated from Eq.~(\ref{eq:ConstitQMassReg})

In our first parameter set, PS1, the calculated current quark mass is $\sim 10$~MeV, larger than the values typically used in the three momentum regularised versions of the model.
As an additional test of sensitivity of the parameters to our fitting procedure we take instead the current quark masses as input, $m_{\ell}=5.5$~MeV and $m_{s}=135.7$~MeV, the pion's mass $m_{\pi}=140$~MeV and its decay constant $f_{\pi}=93$~MeV. By following a similar procedure as above we  determine the other parameters and calculate the constituent quark mass. This leads to a new and substantially different parameter set, PS2, with the constituent quark mass considerably lower. 
When fitting our model parameters, we are enforcing a scale in our model. With this in mind we should compare and choose the parameter set which is both consistent with hadron phenomenology (enforced through the above mentioned fitting procedures) and also favourable for modelling high density quark matter. We will compare the proper time regularised model with the three momentum regularised model~\cite{Masuda:2012ed}.

\subsection{At finite density}
\label{subsec:finden}

At finite density we have conservation of baryon number and associated chemical potentials. To handle this, an extra term is added to our NJL Lagrangian Eq.~(\ref{eq:NJLLag}),
\begin{equation}
\label{eq:LagChem}
\mathcal{L}_{\rm NJL} \rightarrow \mathcal{L}_{\rm NJL} + \bar{\psi} \hat{\mu}\gamma^{0}\psi\quad ,
\end{equation}
where $\hat{\mu}$ is the chemical potential matrix given by $\hat{\mu}=$diag$(\mu_{u}$,$\mu_{d}$,$\mu_{s})$.

The inverse quark propagators in momentum space are now of the form
\begin{equation}
\label{eq:InvPropFD}
S^{-1}_{i}(p) = ( p_{0} + \tilde{\mu}_{i})\gamma^{0} - \vec{p}\cdot\vec{\gamma} - M_{i}  
\end{equation}
for each flavour $i$, where we have defined the reduced chemical potential
\begin{equation}
\tilde{\mu}_{i} =  \mu_{i} - 4G_{\rm V}\rho^{\rm v}_{i} \equiv \mu_{i} - 4G_{\rm V}\langle \psi^{\dagger}_{i}\psi_{i}\rangle\quad .
\end{equation}
Using standard methods~\cite{Ripka:1997zb,Swanson:1992cz} the effective potential evaluated at finite density in the MFA is
\begin{eqnarray}
\label{eq:MFEffPot}
&&\hspace*{-0.85cm} V^{\rm NJL}_{\rm MF}(\left\lbrace M_{i}\right\rbrace,\left\lbrace \mu_{i}\right\rbrace) \nonumber\\
  & = & 2iN_{c}  \sum_{i\in \left\lbrace u,d,s\right\rbrace} \int \frac{d^{4}k}{(2\pi)^{4}} \  {\rm Log} \left[ \frac{k^{2}-M^{2}_{i}+i\epsilon}{k^{2}-M_{i0}^{2}+i\epsilon}\right] \nonumber\\
&&+ \sum_{i\in \left\lbrace u,d,s\right\rbrace} \frac{(M_{i}-m_{i})^{2}}{8G_{\rm S}}  - \sum_{i\in \left\lbrace u,d,s\right\rbrace} \frac{(M_{i0}-m_{i})^{2}}{8G_{\rm S}} \nonumber\\
&& -\ 2N_{\rm C} \sum_{i\in\left\lbrace u,d,s\right\rbrace } \int \frac{d^{3}p}{(2\pi)^{3}}\ \Theta (\tilde{\mu}_{i} - E_{p,i}) (\tilde{\mu}_{i} - E_{p,i})\nonumber\\
&& -  \sum_{i\in \left\lbrace u,d,s \right\rbrace }  \frac{(\tilde{\mu}_{i}-\mu_{i})^{2}}{8G_{\rm V}} \quad ,
\end{eqnarray}
where $M_{i0}$ is the vacuum value of the constituent quark mass and $E_{p,i}=\sqrt{\vec{p}^{\ 2}+M_{i}^{2}}$ for flavour $i$. In Eq.~(\ref{eq:MFEffPot}) we have subtracted a constant defining the effective potential, and hence the pressure ($P=-V$), to be zero in vacuum. With this definition of the effective potential and by the Gibbs-Duhem relation, the energy density also vanishes in vacuum. This is a common definition of the model as it is only defined up to a constant. However, some authors choose to exploit this degree of freedom by introducing a ``Bag" constant, allowing the model to have non-zero values for the above thermodynamic variables in vacuum. 
The first two lines of Eq.~(\ref{eq:MFEffPot}) contain terms which are divergent and must be regularised. 
We choose to regularise using Schwinger's covariant proper time method~\cite{Klevansky:1992qe}. The choice of the regularisation procedure is a defining decision of any NJL model.

After the analytic continuation to Euclidean space the stationary condition for the path integral translates to the condition that the mean field effective  potential is determined at a global minimum and must therefore satisfy
\begin{eqnarray}
\frac{\partial V^{\rm NJL}_{\rm MF}}{\partial M_{i}}=0
& 
\quad 
{\rm and } 
\quad
&
\frac{\partial^{2} V^{\rm NJL}_{\rm MF}}{\partial M^{2}_{i}}\geq 0 
\quad 
\forall i\in\left\lbrace u,d,s\right\rbrace 
\quad\quad .
\end{eqnarray}
The gap equation therefore acquires an additional contribution at finite density, which acts to restore chiral symmetry.

\subsection{Flavour independent vector interaction}
\label{subsec:flavIndepV}

We anticipate that the vector interaction is important (as is well known, see Ref.~\cite{Klahn:2006iw}) and that the strength and type of this interaction is crucial for a realistic description of quark matter. For this reason we introduce an alternative ``simplified" vector interaction which is flavour independent, such that
\begin{equation}
\mathcal{L}_{\rm v} = -g_{\rm V} (\bar{\psi}\gamma^{\mu}\psi)^{2} \quad .
\label{eq:FlavIndepInt}
\end{equation}
This form of vector interaction has been used in many NJL studies. Particularly interesting are those that use it to produce high mass neutron stars if the coupling is large enough, see for example Refs.~\cite{Masuda:2012ed,PhysRevC.90.045801}.
With the vector interaction given by Eq.~(\ref{eq:FlavIndepInt}), rather than the flavour dependent interaction in Eq.~(\ref{eq:NJLLag}), both the reduction in the chemical potentials and the contribution to the effective potential must be recalculated.

In symmetric two flavour quark matter ($\rho_{\rm u} =\rho_{\rm d}$ and $\rho_{\rm s}=0$) these two interactions are equivalent but differ otherwise. In asymmetric two flavour quark matter they will differ and there should be a substantial difference when strange quarks are present.
In three flavour symmetric quark matter ($\rho_{\rm u} = \rho_{\rm d} =\rho_{\rm s}$) the additional cross terms for the flavour independent interaction could give a substantial increase in  pressure coming from the vector contribution.
Of course, each of the quark chemical potentials will be reduced by the same amount, determined by the total quark density, as opposed to the flavour dependent interaction, where each quark's chemical potential is only reduced by its  own density. 
 Consequently, the type of vector interaction could be important in the description of hybrid and quark stars, particularly when strange quarks are involved.

\subsection{Quarks in beta-equilibrium}
\label{subsec:QBE}

Thermal equilibrium of quarks and leptons with respect to the weak and strong interactions, under the constraints of charge and baryon number conservation, is described by the following system of equations:
\begin{eqnarray}
\frac{2}{3}\rho_{\rm u}^{\rm v} -\frac{1}{3}\left( \rho_{\rm d}^{\rm v} + \rho_{\rm s}^{\rm v}\right) -\rho_{\rm e}^{\rm v} -\rho_{\mu}^{\rm v} & = & 0\label{eq:Q}\\
\rho -\frac{1}{3}\left( \rho_{\rm u}^{\rm v} +\rho_{\rm d}^{\rm v} + \rho_{\rm s}^{\rm v}\right)  & = & 0\label{eq:B}\\
 \mu_{\rm d} -\mu_{\rm u} -\mu_{\rm e} &=& 0\label{eq:W1}\\
 \mu_{\rm d} -\mu_{\rm s} & = & 0\label{eq:W2}\\
 \mu_{\mu}-\mu_{\rm e} & = & 0 \label{eq:W3}\quad .
\end{eqnarray}
The relations imposed in Eqs.~(\ref{eq:W1}--\ref{eq:W2}) are between the thermodynamic chemical potentials and not the reduced chemical potentials.
In terms of the individual quark densities ($\rho_{\rm i}^{\rm v}$), the total quark density ($\rho_{\rm tot}$) and the total baryonic density ($\rho$) are defined as
\begin{equation}
\rho \equiv  \frac{\rho_{\rm tot}}{3} \equiv \frac{1}{3}\sum_{i\in\left\lbrace u,d,s\right\rbrace }\rho_{\rm i}^{\rm v} \quad .
\end{equation}
 In the limit of zero vector coupling, the individual number density of each quark species is related to the respective chemical potential by
 \begin{equation}
 \label{eq:Chem_Den}
 \rho_{\rm i}^{\rm v} =\frac{ ( p_{\rm F}^{i} )^{3}}{\pi^{2}} = \frac{(-M_{i}^{2}+\mu_{i}^{2})^{3/2}}{\pi^{2}} 
  \xrightarrow{G_{\rm V}\neq 0}\frac{(-M_{i}^{2}+\tilde{\mu}_{i}^{2})^{3/2}}{\pi^{2}}  \quad ,
 \end{equation}
where $p_{\rm F}^{i}$ is the quark Fermi momentum. 
For non-zero vector coupling, the chemical potential in Eq.~(\ref{eq:Chem_Den}) is replaced with its reduced counterpart.
The lepton chemical potentials are once again given by Eq.~(\ref{eq:chempots}). 
This system of equations combined with the three gap equations, is then solved to determine the particle content and thermodynamic behaviour of three flavour quark matter in beta-equilibrium with leptons.

The pressure of quark matter is calculated from the thermodynamic relation 
\begin{equation}
P = -V_{\rm total} = -V^{\rm NJL}_{\rm MF}\left( \left\lbrace M_{i}\right\rbrace, \left\lbrace \mu_{i} \right\rbrace \right)  - V_{l}(\left\lbrace \mu_{l}\right\rbrace ) \quad ,
\end{equation}
where $V_{l}$ is the effective potential contribution of the non-interacting leptons. This gives the same pressure contribution as in Sec.~\ref{sec:hadronic}. 
The energy density is obtained from the following formula 
\begin{equation}
\epsilon_{\rm total} = V_{\rm total} + \sum_{i\in \left\lbrace u,d,s,e,\mu \right\rbrace }\mu_{i}\rho^{\rm v}_{i} \quad ,
\end{equation}
where in the second term $\mu_{i}$ is the un-reduced thermodynamic chemical potential.

\subsection{Quark matter numerical results}
\label{subsec:Quark_NumRes}
\begin{figure*}
\includegraphics[width=0.9\textwidth]{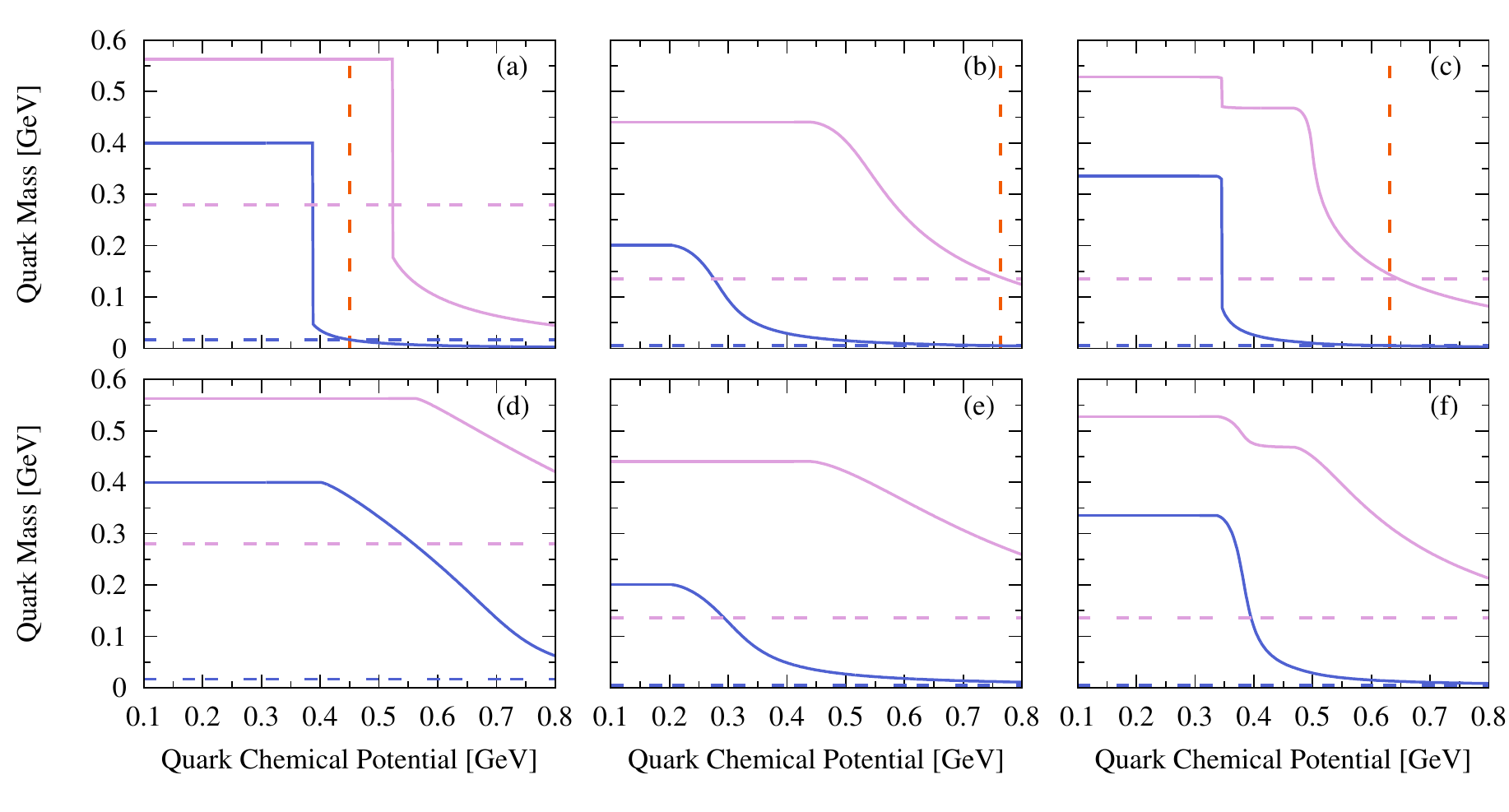} 
\caption{Constituent quark masses of light (blue lines) and strange (purple lines) quarks as a function of chemical potential ($\mu=\mu_{\ell}=\mu_{s}$) for the parameter set (a,d) PS1, (b,e) PS2 and (c,f) HK. The constant light (blue) and strange (purple) current quark masses are shown by the horizontal dashed lines. In the first row the vector coupling is set to zero, and in the second it is $G_{\rm V}=G_{\rm S}$. The vertical red dashed line in the top row indicates the critical chemical potential.}
\label{fig:Q_MassesVsMu}
\end{figure*}

In this section we will present and discuss the numerical results of the NJL model parameter sets 
PS1, PS2 and HK. We begin with the behaviour of the constituent quark mass as a function of quark chemical potential. This is 
followed by the properties of quark matter in beta equilibrium with leptons. 

Chiral symmetry is broken explicitly in all parameter sets considered (see Table~\ref{table:NJLPSets}) by the presence of a non-zero current quark mass but it is also broken dynamically in vacuum for sufficiently large coupling strength~\cite{Vogl:1991qt}. All parameter sets produce a constituent quark mass which is significantly larger than the current quark mass in vacuum for all flavours.
The constituent quark masses, or equivalently their condensates, are the order parameters of chiral
symmetry.  The chirally broken phase is marked by a large constituent quark mass and its approximate
restoration is expected to occur at large chemical potential for all three parameter sets.

The numerical results presented in Figure~\ref{fig:Q_MassesVsMu}, show the 
behaviour of the constituent quark masses, as a function of the quark chemical potential ($\mu = \mu_{\ell}=\mu_{s}$) for each parameter set. In the first row the vector coupling is set to zero and $G_{\rm V} = G_{\rm S}$ in the second.

In the NJL model, which models the dynamical generation of mass breaking chiral symmetry, it is unnatural for the constituent quark mass to be smaller than the current quark mass. Obviously, this can occur when the finite density terms overwhelm the vacuum terms in the gap equation. One would naively expect that this would not occur before the UV cut-off, which was introduced to regulate the model, effectively setting the scale of the model using relevant hadron phenomenology. This is quite plainly not the case for parameter PS1, see Fig.~\ref{fig:Q_MassesVsMu} plot (a). The constituent quark mass for light quarks very abruptly drops at $\mu\simeq 388$~MeV. Shortly after this first order transition occurs, the constituent quark mass becomes smaller than the current quark mass at $\mu\simeq 451$~MeV\textemdash which is moderately lower than the cut-off $\Lambda_{\rm UV} \simeq 637$~MeV. More serious, however, is the behaviour of the constituent strange quark mass, which drops sharply below its current quark mass as soon as it is energetically favourable to appear at $\mu\simeq 524$~MeV.

The behaviour of the constituent quark masses as a function of quark chemical potential for parameter set PS2 are markedly different from the behaviour for PS1, as shown in Fig.~\ref{fig:Q_MassesVsMu}. To begin with, there is no-longer a first order transition for all flavours in the absence of vector coupling. The transition between the chirally broken phase and the symmetric phase is smooth, with the constituent quark masses still going below the current quark masses at $\mu\simeq 763$~MeV for light quarks and $\mu\simeq 771$~MeV for strange quarks\textemdash once again at a chemical potential moderately lower than the cut-off $\Lambda_{\rm UV}=1.0789$~GeV.

The three-momentum regularised NJL model with t'~Hooft determinantal term has been used extensively in the literature~~\cite{Vogl:1991qt,Klevansky:1992qe,Hatsuda:1994pi,Buballa:2003qv}, and in particular, it was recently used to study hybrid stars in Ref.~\cite{Masuda:2012ed}. This variation of the NJL model is included for comparative purposes using the HK parameter set~\cite{Hatsuda:1994pi}. Table~\ref{table:NJLPSets} contains the HK parameter set for convenience, but the interested reader is referred to Ref.~\cite{Hatsuda:1994pi} for how it was obtained. The values of the current quark masses in this parameter set were used as input for our proper time regularised parameter set PS2, so a close comparison could be performed.
Current quark masses of the HK parameter set are the same as in PS2, but the constituent quark masses are considerably larger and comparable to the ones used in PS1. Moreover, the scalar coupling and UV cut-off are comparable to their counter parts in PS2 and PS1, respectively.

The 't~Hooft determinantal term has a pronounced effect on the constituent quark masses\textemdash see Fig.~\ref{fig:Q_MassesVsMu} plots (c,f). The mixing of different flavours produced by the flavour determinant means that the constituent quark masses of different flavours are inter-related. In plots (c,f) the strange quark mass drops abruptly when the light quark mass drops, before it is actually favourable to appear, illustrating the dependence of the strange quark mass on the light quark condensates.

Of considerable interest is that, unlike in the proper time regularised models described above, the constituent quark masses of both the light and strange quarks do not go below the current quark mass until the quark chemical potential approximately reaches the UV cut-off. This is somewhat disconcerting as we could interpret this as a signal that the NJL model with our chosen regularisation scheme is breaking down and, in the case of PS1, it is occurring at only a moderate chemical potential. In the case of PS2 it breaks down at a chemical potential greater than in the HK parameter set, but still lower than the UV cut-off.
As the regularisation scheme is a defining feature of the model, this difference should be clearly understood.
Furthermore, we will be applying the NJL model to describe hybrid stars in Sec.~\ref{sec:crossover}, where the inner core densities are expected to be immense. The large quark chemical potentials that are anticipated to be achieved could surpass the breaking point of the model.

As pointed out in appendix B of Ref.~\cite{Maedan:2009yi}, this can be understood by studying the in-medium gap equation in detail.
The in-medium gap equation takes the form of a sum of vacuum and finite density contributions.
When the finite density contribution is greater than the vacuum contribution, the constituent quark mass is smaller than the current quark mass. This occurs at some critical chemical potential denoted $\mu_{\rm crit}$. The value of the critical chemical potential is dependent on the regularisation scheme through the vacuum contribution to the  gap equation only, as the Fermi term is finite~\footnote{However, this too could be regularised.}. It can be estimated by performing an expansion of vacuum and finite density terms about $M_{i}/\Lambda_{\rm UV} = 0$. In the proper time scheme, the pertinent integral that must be expanded needs be treated with care, because of the singular nature of this integral in the limit $M_{i}/\Lambda_{\rm UV}\to 0^{+}$. Specifically, it contains a logarithmic singularity. Because of this singularity the Taylor series expansion has a zero radius of convergence and it therefore must be treated as an asymptotic expansion rather than a Taylor expansion. 
The critical chemical potential to the lowest order is then
\begin{equation}
\mu_{\rm crit} \simeq \frac{\Lambda_{\rm UV}}{\sqrt{2}}
\simeq 
\left\lbrace 
\begin{array}{ccc}
450{\rm ~MeV} & \quad & {\rm for\ PS1} \\
763{\rm ~MeV} & \quad & {\rm for\ PS2} 
\end{array} 
\right.
 \quad .
\end{equation}
Similarly, one can show in the three-momentum cut-off regularisation that
$\mu_{\rm crit} \simeq \Lambda_{\rm UV}$~\cite{Maedan:2009yi}.

The vector interaction renormalises the chemical potential, effectively increasing it. However, it is the reduced chemical potential, $\tilde{\mu_{i}}$, that appears, for example, in the in-medium gap equation. The critical value of the chemical potential derived above will now apply to the reduced chemical potential. It will then provide a limiting value to the reduced chemical potential, up to which we can consider the model to be reliable in the presence of the vector interaction.

We allow the vector coupling to vary from zero to up to being equal to the scalar coupling, in order to understand its affect on the model. 
On increasing the vector coupling in each of the models the transition from the chirally broken phase to the symmetric phase occurs more smoothly. In particular, in the case of parameter set PS1 it changes the transition from a first order transition to a second order transition. 
This is not surprising as this effect has been seen in studies of two-flavour quark matter~\cite{Buballa:2003qv} using a flavour independent vector interaction. It has also been seen in NJL model studies of the QCD phase diagram in the $T$\textendash $\mu$ plane, whereby the vector interaction shifts the critical point closer to the $\mu$ axis~\cite{Bratovic:2012qs,Hell:2012da,Friesen:2014mha}.
For each of the models, the vector coupling defers chiral restoration to a larger chemical potential. The reduction of the chemical potential from the vector interaction curtails the effect of the finite density contribution to the gap equation, meaning that the constituent quark masses approach their current quark masses at greater chemical potential.

Based on the above discussions, the PS2 parameter set will likely make a more reliable description of hybrid stars.  
In the modelling of quark matter in beta-equilibrium with leptons, we will restrict ourselves to the parameter sets PS2 and HK.

%Based on the above discussions, the PS1 parameter set will not likely make a reliable description of hybrid stars.  
%In the modelling of quark matter in beta-equilibrium with leptons, we will restrict ourselves to the parameter sets PS2 and HK.

\begin{figure}%[H]
%\centering
\includegraphics[width=0.5\textwidth]{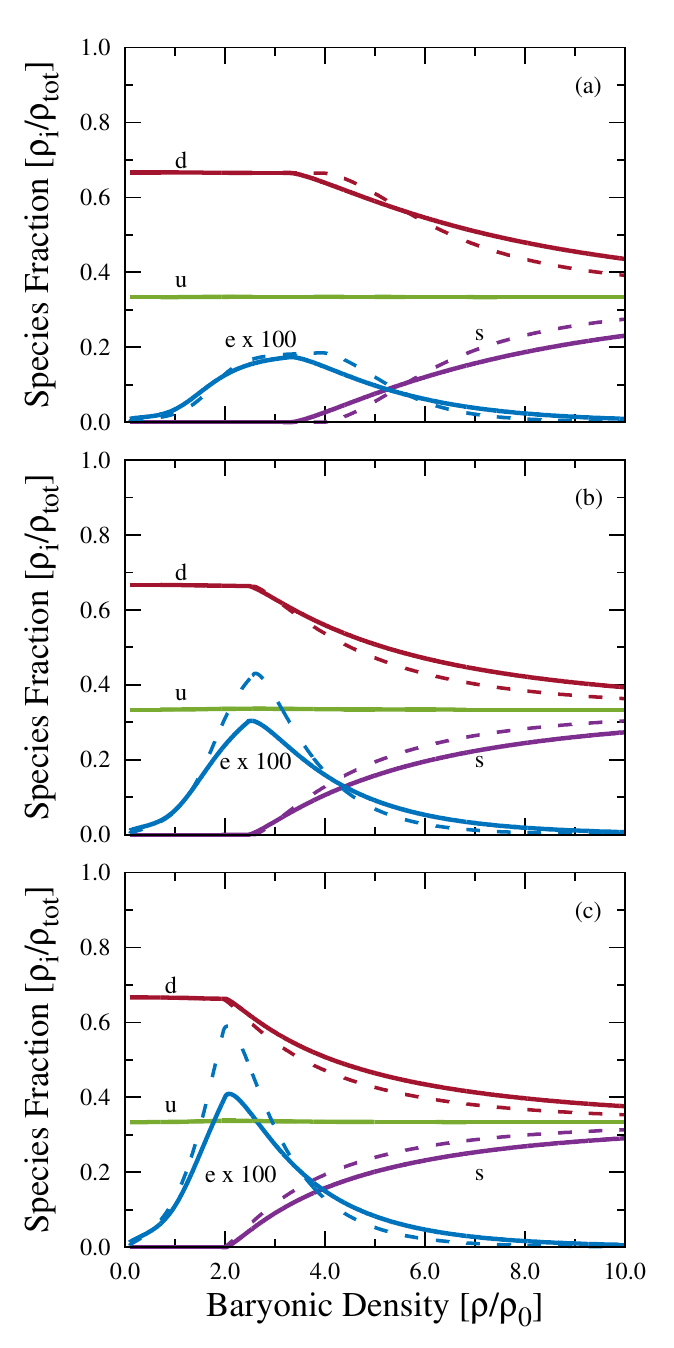} 
\caption{
Species fractions as a function of density for beta-equilibrium quark matter for parameter set PS2 (solid) and HK (dashed). Each of the particle number densities is divided by the total quark density $\rho_{\rm tot} =\rho_{d}+\rho_{u}+\rho_{s} = 3\rho$. The down quark fraction is red, up green, strange purple. The electron fraction (blue) is multiplied by 100 so as to be visible on the same plot. Note that the electron fraction defined here differs by a factor of $1/3$ from the figures in Sec.~\ref{sec:hadronic}. Plot (a) zero vector coupling and non-zero flavour independent vector coupling (b) flavour dependent vector interaction with $G_{\rm V} = G_{\rm S}/2$ and (c) flavour dependent vector interaction with $G_{\rm V} = G_{\rm S}$. Here we use the saturation density $\rho_{0} = 0.17$~fm$^{-3}$. }
\label{fig:BetaEq_SpFr}
\end{figure}

\begin{figure*}%[H]
%\centering
\includegraphics[width=0.9\textwidth]{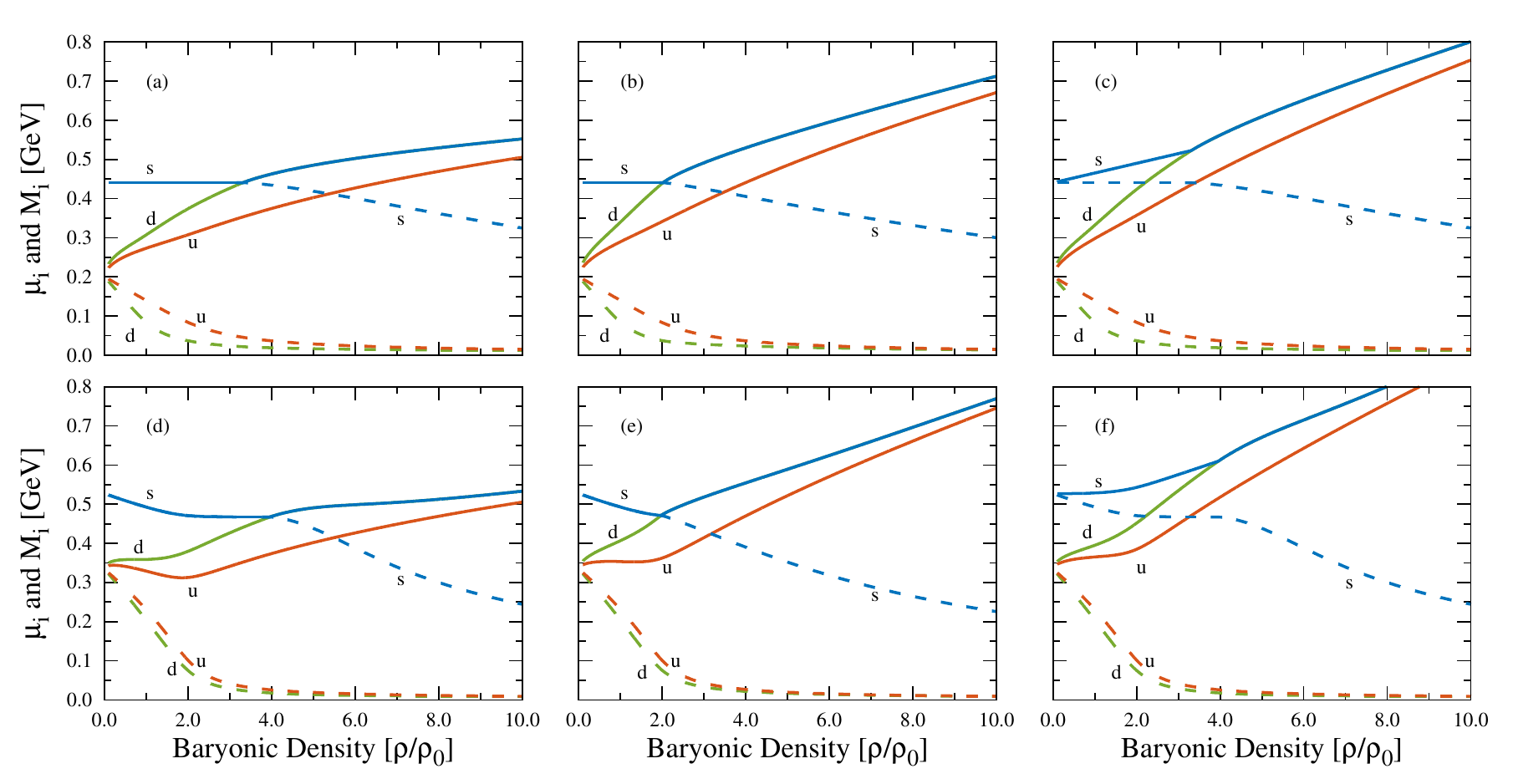} 
\caption{Chemical potentials (solid) and constituent quark masses (dashed) as a function of total baryonic density ($\rho$) for both flavour independent and flavour dependent vector interactions.
The line colours for the quarks are up (orange), down (green) and strange (blue). 
Plots (a--c) PS2 model with $G_{\rm V} = 0, G_{\rm S}$ and $g_{\rm V} = G_{\rm S}$, respectively. Plots (d--f) HK model with $G_{\rm V} = 0, G_{\rm S}$ and $g_{\rm V} = G_{\rm S}$, respectively. 
Here we use the saturation density $\rho_{0} = 0.17$~fm$^{-3}$. }
\label{fig:ChemsAndMasses_Combined}
\end{figure*}

Figure~\ref{fig:BetaEq_SpFr} shows the species fractions as a function of total baryon density. In this figure, the results for both the PS2 and HK parameter sets are shown and are found to have similar particle content and behaviour.
In contrast to hadronic calculations, the only leptons to appear are electrons and in a reduced number. In Fig.~\ref{fig:BetaEq_SpFr}, the electron fraction was multiplied by 100 to make it clearly visible on the same plot as the quark fractions.  
The species fraction in the absence of a vector interaction is shown in Fig.~\ref{fig:BetaEq_SpFr}~(a). Incorporating a non-zero flavour independent vector interaction leaves the number densities of the particles unchanged. Similar plots showing species fractions in the three flavour NJL model with flavour independent vector interaction can be found in Refs.~\cite{Masuda:2012ed} and \cite{PhysRevC.90.045801}.
The onset of strangeness occurs at a slightly lower density using the  PS2 model than in the HK model.
The strange quarks appear at $\rho\simeq 3.32\rho_{0}$ in PS2 and at $\rho \simeq 3.98\rho_{0}$ in HK.
As expected, the appearance of strange quarks reduces the number of down quarks because of the charge neutrality constraint and the up quark fraction remains approximately constant over the density range considered. The strength of the flavour independent vector interaction does not change the species fraction as a function of density, but does have an affect on the thermodynamic variables, as will be discussed below.
However, in the case of a flavour dependent vector interaction, particle densities do vary with vector coupling strength, see plot (b--c) of Fig.~\ref{fig:BetaEq_SpFr}. Figure~\ref{fig:BetaEq_SpFr}~(b--c) shows the onset of strangeness occurs at lower density with increasing strength of the flavour dependent vector interaction for both PS2 and HK models. With varying the vector coupling between $(0,G_{\rm S})$, the threshold density for strange quarks is in the range $\sim 2\rho_{0}$--$4\rho_{0}$.

Figure~\ref{fig:ChemsAndMasses_Combined}  shows chemical potentials and constituent quark masses as a function of total baryonic density and  illustrates the affect of flavour dependent and independent vector interactions.
 Both the PS2 and HK models show similar trends, with the exception that in the HK parameter set there is more curvature of the chemical potentials at low density. This is undoubtedly connected to the t'~Hooft term causing the condensates to be dependent on one another. 
In the HK parameter set, the strange quark mass decreases as the light quark masses decrease\textemdash even when strange quarks have not yet appeared. On the other hand, in PS2 (with a flavour dependent vector interaction) it remains constant until it is energetically favourable to be produced. This is because the quark condensates of each flavour are independent of each other in PS2.
From Fig.~\ref{fig:BetaEq_SpFr}, we see that at low density only the light quarks are present for both parameter sets and all variations of the vector interaction considered. 
For models incorporating the flavour dependent vector interaction, we have $\mu_{s} = M_{s}$ at zero strange quark density, which can be seen in cases (b,e) of Fig.~\ref{fig:ChemsAndMasses_Combined}. 
As the baryonic density increases, the separation of the chemical potential curve from the constituent quark mass curve can be clearly seen for the strange quark. This coincides with the appearance of strange quarks. With increasing strength of the vector interaction, the down quark chemical potential increases faster with increasing density, leading to an earlier onset of strange quarks\textemdash compare with Fig.~\ref{fig:BetaEq_SpFr}.
However, for models with the flavour independent vector interaction $\mu_{s} = M_{s} + 2g_{\rm V}(\rho_{u}+\rho_{d})$ at zero strange quark density. Before the density threshold is reached for strange quarks, there is a separation between the strange quark's chemical potential and its constituent quark mass owing to the already present light quarks. This can be seen in plots (c,f) of Fig.~\ref{fig:ChemsAndMasses_Combined}.

Moreover, as we are working in the isospin symmetric limit, i.e the current quark masses of the light quarks are equal, the only cause for a difference between the light constituent quark masses is the finite density contribution. In contrast to the usual free space situation, where the up quark is lighter than the down quark and thereby the neutron heavier than the proton, in-medium the down quark is found to be lighter than the up quark. This is because of the conditions of beta-equilibrium under charge and baryon number conservation. The down quark fraction is greater than the up quark fraction, as can be seen in Fig.~\ref{fig:BetaEq_SpFr} and so its effective mass is reduced more than that of the up quark. However, this difference is small and decreases with increasing density. 

The reason for the differences between the flavour dependent and independent vector interactions lies in the imposed beta-equilibrium relations in Eqs.~(\ref{eq:W1}-\ref{eq:W2}). Expressing Eqs.~(\ref{eq:W1}-\ref{eq:W2}) in terms of the reduced chemical potentials for the flavour independent vector interaction, one finds
an equivalence between the beta-equilibrium relations written in terms of the chemical potentials and the reduced chemical potentials. This is the reason why the species fraction of particles do not change with increasing vector coupling. The particle number densities are directly related to the reduced chemical potentials via Eq.~(\ref{eq:Chem_Den}). This, combined with the gap equation for the quark masses, which is also only dependent on the reduced chemical potential, means that the whole system of beta-equilibrium equations is independent of the vector coupling. At a given density the chemical potentials are larger with increasing vector coupling but their increase is cancelled in the beta-equilibrium relations (Eqs.~(\ref{eq:W1}--\ref{eq:W2})) leaving the particle number densities and hence also the constituent quark masses invariant. 

For the models including a flavour dependent vector interaction, there are extra terms which do not cancel and are proportional to the vector coupling. 
The equilibrium conditions do not simplify to relations of the same form between the reduced chemical potentials. Because there remains an explicit dependence on the vector coupling in the beta-equilibrium relations, the particle number densities and hence the constituent quark masses change with variation of the vector coupling.

\begin{figure}
\includegraphics[width=0.45\textwidth]{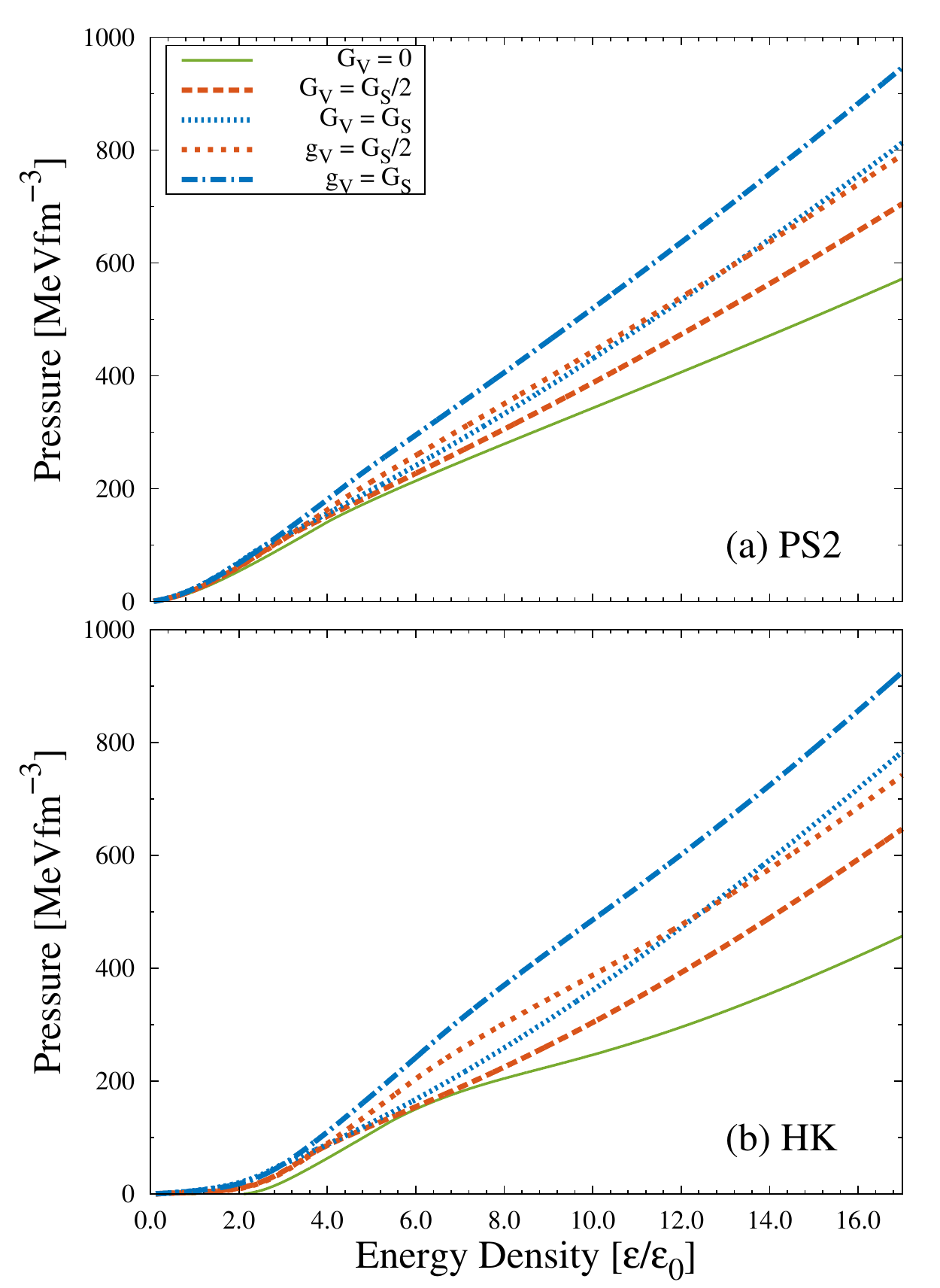} 
\caption{Pressure as a function of energy density for the (a) PS2 and (b) HK parameter sets. Results using the flavour dependent interaction $G_{\rm V}$ (i.e. use Eq.~(\ref{eq:NJLLag}))  and flavour independent interaction $g_{\rm V}$ (i.e. use Eq.~((\ref{eq:FlavIndepInt}))  for different values of the vector coupling. Here we normalise the energy density with $\epsilon_{0} = 140$~MeVfm$^{-3}$.}
\label{fig:PressVsEnDen_Combined}
\end{figure}

Figure~\ref{fig:PressVsEnDen_Combined} plainly shows the appreciable effect the vector interaction has on the pressure as a function of energy density.  Overall, the behaviour of the two parameter sets with change in strength and type of vector interaction is not dissimilar. For both, there is a considerable increase in pressure upon turning on the vector coupling and then increasing it further to be equal to the scalar coupling. As anticipated, the flavour independent vector interaction provides a larger increase in pressure at high density ($\rho\gtrsim 2\rho_{0}$) for both parameter sets. The earlier onset of the strange quark makes the models with a flavour dependent vector interaction a little softer again. However, the vector interaction still produces a stiffer EoS state on increasing the vector coupling.
The PS2 parameter set produces a slightly stiffer EoS, particularly at low density.

\section{Faux crossover construction}
\label{sec:crossover}

In this section we consider a transition from hadronic matter (modelled using the HF-QMC model) to quark matter (using the NJL model). We investigate the possibility of a smooth crossover transition in a purely phenomenological way, whereby we interpolate between the hadronic and quark model equations of state (EoS).

%Maxwell
In sections~\ref{sec:gbem} and \ref{sec:quark}, we considered each phase to be in beta-equilibrium and also charge neutral. The requirement of charge neutrality effectively reduced each phase to a one component system controlled by the baryonic density or equivalently a baryonic chemical potential. 
Built on this foundation, one is naturally led to consider phase transitions in neutron stars modelled assuming a one component description, i.e., a Maxwell construction. This is the simplest possibility for constructing a phase transition and historically the most studied.  The transition point in the Maxwell construction is identified by the conditions of thermal, mechanical and one component chemical equilibrium.
 This first order transition corresponds to a kink in the pressure versus neutron chemical potential plane and a constant pressure plateau in the pressure versus density plane. This plateau connects the hadronic phase to the quark phase.  
 With this sudden jump in the density at constant pressure, the Maxwell construction does not allow for the possibility of a mixed phase where both hadrons and quarks can coexist together. For actual hybrid stars, in this construction, one finds a hadronic outer layer and a dense quark core, with no possibility for a mixed phase in between.

%Gibbs
When modelling phase transitions in neutron stars using the Maxwell construction each phase is considered independently charge neutral. However, as was first pointed out by Glendenning~\cite{Glendenning:1992vb}, if a mixed phase exists then charge neutrality can be achieved globally rather than locally.
To consider this possibility we are led to the Gibbs construction for a multi-component system. In the context of hybrid stars, we have a two component system corresponding to two conserved quantities, namely baryon number and charge.

The Gibbs construction for a first order transition requires that thermal, mechanical and chemical equilibrium are implemented in the mixed phase region.
Chemical equilibrium requires that the, now two, independent chemical potentials (the neutron and electron chemical potentials) of the two oppositely charged phases are equal. Outside of the coexistence region the phase with the greatest pressure is the persistent phase.
As the hadronic phase is known to be the dominant phase at low density, one calculates the hadronic phase and uses the calculated neutron and electron chemical potentials as input into the quark phase calculation of the equation of state. 
If at some density the pressures of the two phases become equal, then a mixed phase is possible. If this mixed phase exists, it can potentially persist over a range of densities with the pressure increasing monotonically with density.
In the mixed phase, hadronic matter will possess a charge and quark matter the opposite charge.

The Maxwell and Gibbs constructions described above are bulk constructions, treating both hadron and quark phases as uniform matter. Important finite size effects are neglected, such as the surface tension at the hadron-quark interface and also the Coulomb interaction. These effects have been shown to lead to the creation of geometrical structures forming phases commonly referred to as pasta or structured mixed phases. From more sophisticated calculations which take these effects into account, see for example Refs.~\cite{Heiselberg:1992dx,Glendenning:1995rd,Glendenning:2001pe,Endo:2005va,Maruyama:2005tb,Maruyama:2005vb,Maruyama2007,Logoteta2013},  it is known that these structures tend to smooth the pressure plateau seen in the Maxwell construction. Moreover, the Maxwell construction can be considered a limiting scenario where the surface tension is large and conversely for the Gibbs construction where it is taken to vanish.

These constructions describing first order transitions typically make it difficult to construct stiff hybrid EoS compatible with large neutron star mass observations, unless the hadronic EoS is already sufficiently stiff. However, some models have been able to produce massive hybrid stars compatible with observation, see for example~\cite{Bonanno2012,PhysRevD.88.085001,Benic2014,PhysRevC.90.045801,Miyatsu2015}. 
The EoS are softer because
 in order to implement the Maxwell and Gibbs constructions, the quark pressure must be less than the hadronic pressure at low neutron chemical potential, intersect at some point, and then remain above with increasing chemical potential. This requirement implicitly restricts the possible hybrid EoS to be softer than hadronic EoS in general.

Moreover, since no known model has a realistic description of the confinement mechanism, this adds to the difficulty in providing a reasonable description of the matter in the transition region.
Model derived hadronic and quark EoS may only provide adequate explanations of strongly interacting matter in the low and the high density limits, respectively. These models may, in fact, be unreliable in the intermediate transition region where the requirements of thermal, mechanical and chemical equilibrium are imposed. 
The requirement of mechanical equilibrium ($P_{\rm Q}=P_{\rm H}$) deserves special emphasis, because models not including confinement would necessarily produce unnaturally large pressure. In some situations, to ensure a transition at a reasonable density, model parameters must be restricted or a bag constant introduced to lower the pressure~\cite{Kojo2015}. Either of these choices will also affect the high density behaviour of the EoS~\cite{Kojo2015}. We use the usual convention, where the pressure and the energy density vanish in vacuum. However, a bag constant could be introduced to produce non-zero values in vacuum. 
For this reason, the Maxwell and Gibbs constructions could fail to capture essential features of the transition region accurately despite the models being otherwise reliable in their respective asymptotic limits.

Moreover, in searching for the hadron\textendash quark phase transition by the Maxwell and Gibbs constructions, the implicit assumption is made that the transition is first order. This is generally assumed, but the hadron\textendash quark transition may not be a first order transition in the interior of the QCD phase diagram. It may take the form of a crossover transition similar to what is predicted by lattice QCD at low density and high temperatures~\cite{PhysRevD.71.034504,Aoki:2006we,PhysRevD.74.054507,PhysRevD.85.054503}. If deconfinement were to take the form of a crossover, we could parametrise our ignorance of the transition region by phenomenologically interpolating between the hadron and quark EoS. This possibility has recently received much attention from several groups~\cite{Masuda:2012kf,Masuda:2012ed,PhysRevC.90.045801,AlvarezCastillo2014,Kojo2015}.

An argument which suggests the possibility that the transition may be a crossover rather than a phase transition follows from the known extended nature of hadrons. With hadrons being a colour singlet cluster of confined quarks, an inference to be drawn from their nature is that a progressive transition to quark matter may occur where hadrons and quarks coexist and interact with one another. As the densities reached inside neutron stars are generally thought not to be greater than $10\rho_{0}$, the quarks are most likely not asymptotically free, but are rather still strongly interacting~\cite{Masuda:2012ed,Masuda:2012kf}. It is well known that including a vector interaction among quarks can significantly stiffen an EoS~\cite{Klahn:2006iw}, meaning that if a crossover transition to a stiff quark EoS occurred at low enough density, this would therefore offer a means to generate massive neutron stars~\cite{Masuda:2012ed,Masuda:2012kf}.

Hybrid EoS were previously calculated using a Gibbs construction employing a Hartree QMC model and a simpler version of the NJL model without vector interactions in Refs.~\cite{Carroll:2008sv,CarrollsThesis}. More recently, a Gibbs construction was employed between a different variation of the Hartree-Fock QMC model and a bag model~\cite{Miyatsu2015}. 
We will not consider the above constructions further. Instead, we will contemplate the possibility that the transition is actually a smooth crossover. 
This will be done using the Hartree\textendash Fock Quark-Meson Coupling (QMC) model developed in Sec.~\ref{sec:hadronic} to describe  
the hadronic phase and the three flavour NJL model developed in Sec.~\ref{sec:quark} for the quark phase.

\subsection{Interpolation construction}
\label{subsec:InterpConstr}

If we assume that we understand how the low and high density matter behaves asymptotically, then we can parametrise our ignorance of the intermediate region where the phase transition occurs using an interpolating scheme. It should be understood that the choice of interpolating scheme is not unique. 
Masuda {\it et al}~\cite{Masuda:2012ed,Masuda:2012kf} investigated two different interpolation constructions, pressure versus baryonic density and energy density versus baryonic density employing a hyperbolic tangent function. Hell and Weise~\cite{PhysRevC.90.045801} interpolated pressure as a function of energy density using a similar function. 
Alvarez Castillo {\it et al}~\cite{AlvarezCastillo2014} and Kojo {\it et al}~\cite{Kojo2015} interpolated pressure as function of baryonic chemical potential using a gaussian and polynomial description, respectively. For each interpolation method, one thermodynamic variable was interpolated as a function of another, then the remaining variables were calculated from the interpolated variable. This results in additional thermodynamic corrections to the calculated variables beyond mere interpolation. These additional corrections are meant to preserve thermodynamic consistency between the variables, which may be important in applications to physical systems such as hybrid stars.  However, since these corrections originate from a phenomenological interpolation, it is not clear whether they are physically meaningful or simply an artefact of the interpolation construction used. Only a deeper understanding of QCD thermodynamics can answer this. For this reason, we show numerical results with and without this thermodynamic correction.

We follow Ref.~\cite{Masuda:2012ed} and interpolate the energy density as a function of total baryonic density.
To facilitate the transition between the hadron and quark EoS we introduce the following sigmoid interpolating functions
\begin{equation}
f_{\pm}(\rho) =\frac{1}{2}\left( 1\pm \tanh (X)\right)  \quad ,
\label{eq:fpm}
\end{equation}
where $X = \frac{\rho-\bar{\rho}}{\Gamma}$ and the transition region is chosen to be $\rho\in\left[\bar{\rho}-\Gamma,\bar{\rho}+\Gamma \right]$ with $(\bar{\rho},\Gamma)=(3\rho_{0},\rho_{0})$ and $\rho_{0}=0.16$fm$^{-3}$. The transition from hadronic EoS to the quark EoS is centred about $\bar{\rho}$ with the width of the transition region determined by $\Gamma$. These sigmoid functions are continuous, monotonic and differentiable, varying smoothly between the horizontal asymptotes of 0 and 1. 
There is no physical argument for these functions other than we want a smooth function to facilitate the transition from the hadronic EoS to the quark EoS producing a faux crossover transition. 
In this manner, two EoS based on different models, including complementary physics and aimed at describing matter in different density regimes, can be smoothly transitioned between in a reasonable, but phenomenological way.

The energy density is interpolated using Eq.~(\ref{eq:fpm}) by
\begin{eqnarray}
\label{eq:Interpolated_EnDen}
\epsilon (\rho) & = & \epsilon_{\rm HP}(\rho) f_{-}(\rho) + \epsilon_{\rm QP}(\rho)f_{+}(\rho)  \quad .
\end{eqnarray}
Note that the functions $f_{\pm}(\rho)$ cannot be interpreted as the quark or hadronic matter volume fraction (as in a Gibbs construction mixed phase), they merely interpolate the energy density.

When the energy density is taken as the interpolated variable as a function of density, the pressure is then calculated from this interpolated energy density (Eq.~(\ref{eq:Interpolated_EnDen})), using
\begin{equation}
P(\rho) = \rho^{2} \frac{\partial (\epsilon/\rho)}{\partial \rho}\quad .
\end{equation}
This leads to
\begin{equation}
\label{eq:P_DeltaP}
P(\rho)  =  P_{\rm HP}(\rho) f_{-}(\rho) + P_{\rm QP}(\rho)f_{+}(\rho)  + \Delta P \quad ,
\end{equation}
where the correction 
\begin{equation}
\label{eq:DeltaP}
\Delta P = \rho \left[ \epsilon_{\rm QP}(\rho)g_{+}(\rho) +\epsilon_{\rm HP}(\rho)g_{-}(\rho)\right] \quad .
\end{equation}
The functions $g_{\pm}(\rho)$ are the density derivatives of the interpolating functions, 
\begin{equation}
\label{eq:gpm}
g_{\pm}(\rho) = \frac{d f_{\pm}(\rho)}{d\rho} = \pm \frac{2}{\Gamma}(e^{X} + e^{-X})^{-2} \quad .
\end{equation}
Functions defined as the derivative of a sigmoid function are bell shaped curves because of the inherent ``s" shape of all sigmoid functions. The thermodynamic correction,  $\Delta P$, will only contribute significantly in the transition region. It is also dependent on the difference of the energy density between the two EoS. If a narrower transition region is chosen, then the bell curve will be more sharply peaked with a larger maximum producing a more substantial contribution to the pressure. 

Using the above procedure we can easily construct many hybrid EoS.
However, we cannot indiscriminately interpolate between hadronic and quark EoS. Rather we should also impose additional criteria to ensure
 we obtain a physically meaningful EoS. The requirements that the EoS be both stable and causal impose stringent constraints ruling out 
 many possible interpolations.

In interpolating between the two EoS, the requirement of stability, i.e., the pressure gradient be greater than zero, 
\begin{equation}
\frac{dP}{d\rho}>0 \quad ,
\label{eq:stability}
\end{equation}
is very restrictive.
Interpolated EoS that do not meet this requirement are not useful in modelling hybrid stars. It is clear from the interpolating functions and the EoS presented in earlier sections, this method will lead to an interpolated EoS that satisfies this constraint, if we omit the thermodynamic correction to the pressure. However, the additional correction may induce inflection points in the EoS, possibly leading to an instability.

By a causal EoS we simply mean an EoS where the speed of sound in matter, $c_{s}$, is less than the speed of light ($c=1$):
\begin{equation}
c_{s}^{2} = \frac{dP}{d\epsilon} < 1 \quad .
\end{equation}
Here we simply calculate it from a high order polynomial fit to the EoS data file. Besides acting as a constraint, it is also a useful measure of the stiffness of an EoS.

\subsection{Faux crossover numerical results}
\label{subsec:Crossover_NumRes}

Throughout this section the interpolations shown in figures, unless otherwise stated, are between the ``Standard" or baseline scenario of the HF-QMC model and the PS2 and HK models incorporating a flavour dependent vector interaction with the transition region chosen to be $\left( \bar{\rho},\Gamma\right) =\left(3\rho_{0},\rho_{0} \right)$. Variations beyond these constructions are examined in Tables~\ref{table:3rho0} and \ref{table:VaryRhoBar}.

\begin{figure}[ht]
%\centering
\includegraphics[width=0.5\textwidth]{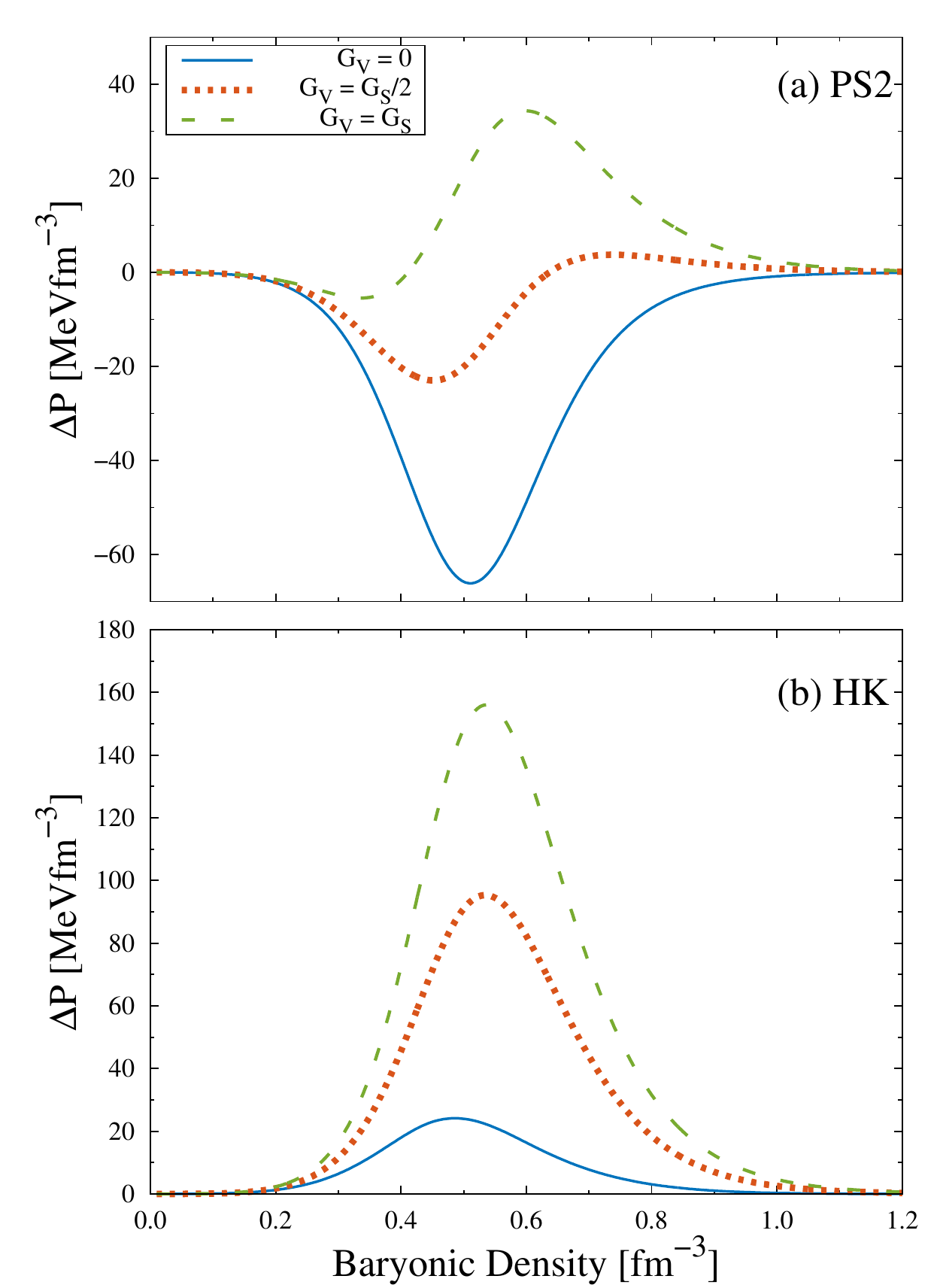} 
\caption{Thermodynamic correction $\Delta P$ as a function of total baryonic density as arising from the interpolation. For plot (a), the interpolation is between the ``Standard" or baseline scenario of the HF-QMC model and the proper time regularised PS2 NJL model with flavour dependent vector interaction. Similarly for plot (b), but with the three momentum regularised NJL model with flavour dependent vector interaction. The crossover region is chosen to be $\left( \bar{\rho},\Gamma\right) =\left( 3\rho_{0},\rho_{0}\right)$. Specific curves for both plots are indicated in the key of plot (a).}
\label{fig:DeltaPVsDen_Multi}
\end{figure}

There is a noticeable difference between the PS2 and HK models, with the HK models producing an energy density greater than the hadronic energy density for all values of the vector coupling. In the case of the PS2 model with no vector interaction the hadronic energy density is greater than the quark energy density for the density range shown. When the vector interaction is increased to half the scalar coupling, the hadronic energy density is greater than the quark energy density up until the density reaches $\rho\sim 0.7$~fm$^{-3}$, then the quark energy density is greater. The difference of the quark and hadronic energy density significantly affects the correction to the pressure to maintain thermodynamic consistency. It also dictates the sign of the correction as seen from  Eq.~(\ref{eq:DeltaP}). A large separation of the quark and hadronic energy density curves indicates a larger correction is needed to maintain thermodynamic consistency\textemdash see Fig.~\ref{fig:DeltaPVsDen_Multi}. 
For all HK models, $\Delta P$ is positive and hence it will stiffen the EoS at the beginning of the transition region and soften it towards the end. The strength of the vector interaction significantly influences the magnitude of $\Delta P$ and on increasing its strength $\Delta P$ is amplified considerably. As for the PS2 models, the sign of $\Delta P$ varies with density and the strength of the vector interaction. In the absence of the vector interaction, it is negative because of the density dependence of the difference of the quark and hadron energy densities. Thus, in contrast to HK models, it will soften the EoS at the beginning of the transition region and stiffen it towards the end.

\begin{figure*}%[ht]
\includegraphics[width=0.9\textwidth]{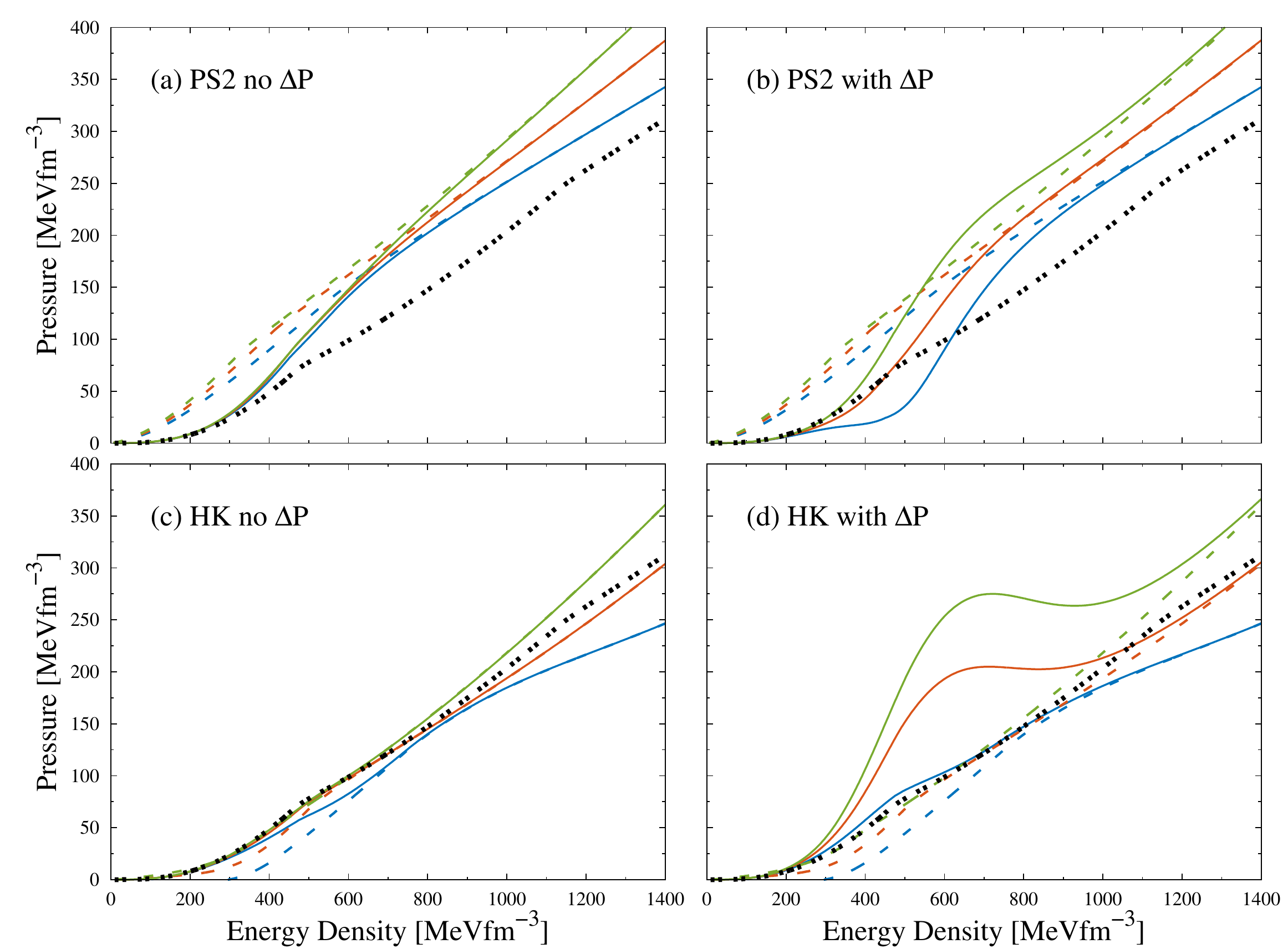} 
\caption{Pressure as a function of energy density. For plots (a,b), the interpolation is between the ``Standard" or baseline scenario of the HF-QMC model and the proper time regularised PS2 model with flavour dependent vector interaction. Similarly for plots (c,d), but with the three momentum regularised model with flavour dependent vector interaction. Plots (a,c) do not include the thermodynamic correction $\Delta P$, whereas plots (b,d) include the correction for thermodynamic consistency. The crossover region is chosen to be $\left( \bar{\rho},\Gamma\right) =\left( 3\rho_{0},\rho_{0}\right)$.  The line type colours indicate the strength of vector coupling used in the quark model, (blue) $G_{\rm V} =0$, (orange) $G_{\rm V} = G_{\rm S}/2$, (green) $G_{\rm V} =G_{\rm S}$. Solid curves are the interpolated EoS and coloured dashed curves the quark EoS. Black dot-dot line is the hadronic EoS.}
\label{fig:PressVsEnDen_Multi}
\end{figure*}

\begin{figure*}%[ht]
\includegraphics[width=0.9\textwidth]{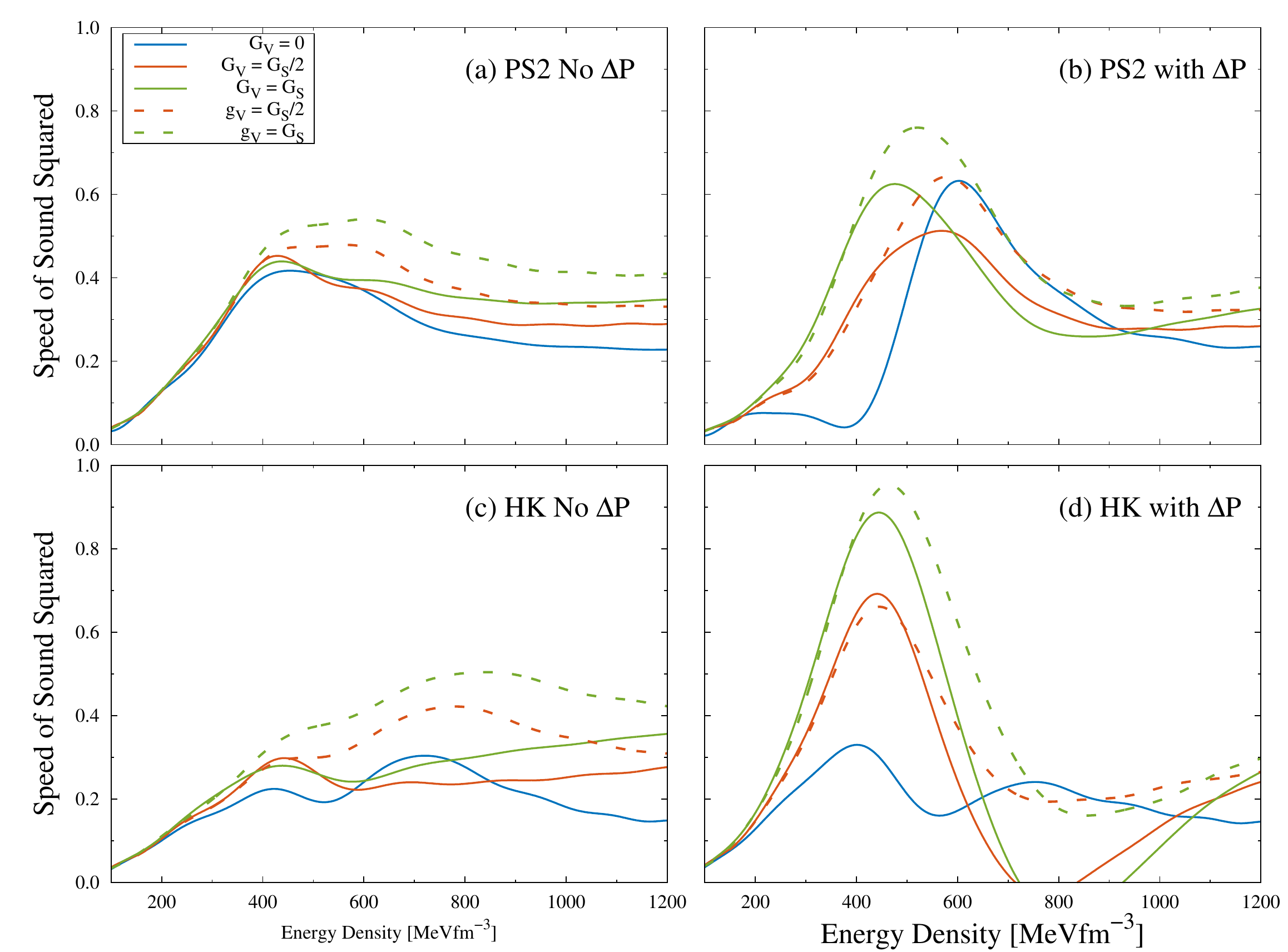} 
\caption{Speed of sound squared (in units of $c$) as a function of energy density. For plots (a,b), the interpolation is between the ``Standard" or baseline scenario of the HF-QMC model and the proper time regularised PS2 model with flavour dependent (solid) and independent (dashed) vector interactions. Similarly for plots (c,d), but with the three momentum regularised model with flavour dependent (solid) and independent (dashed) vector interactions. Plots (a,c) do not include the thermodynamic correction $\Delta P$, whereas plots (b,d) include the correction for thermodynamic consistency. The crossover region is chosen to be $\left( \bar{\rho},\Gamma\right) =\left( 3\rho_{0},\rho_{0}\right)$. Specific curves for all plots are indicated in the key of plot (a). }
\label{fig:Cs2VsEnDen_Multi}
\end{figure*}

\begin{figure*}%[ht]
\includegraphics[width=0.9\textwidth]{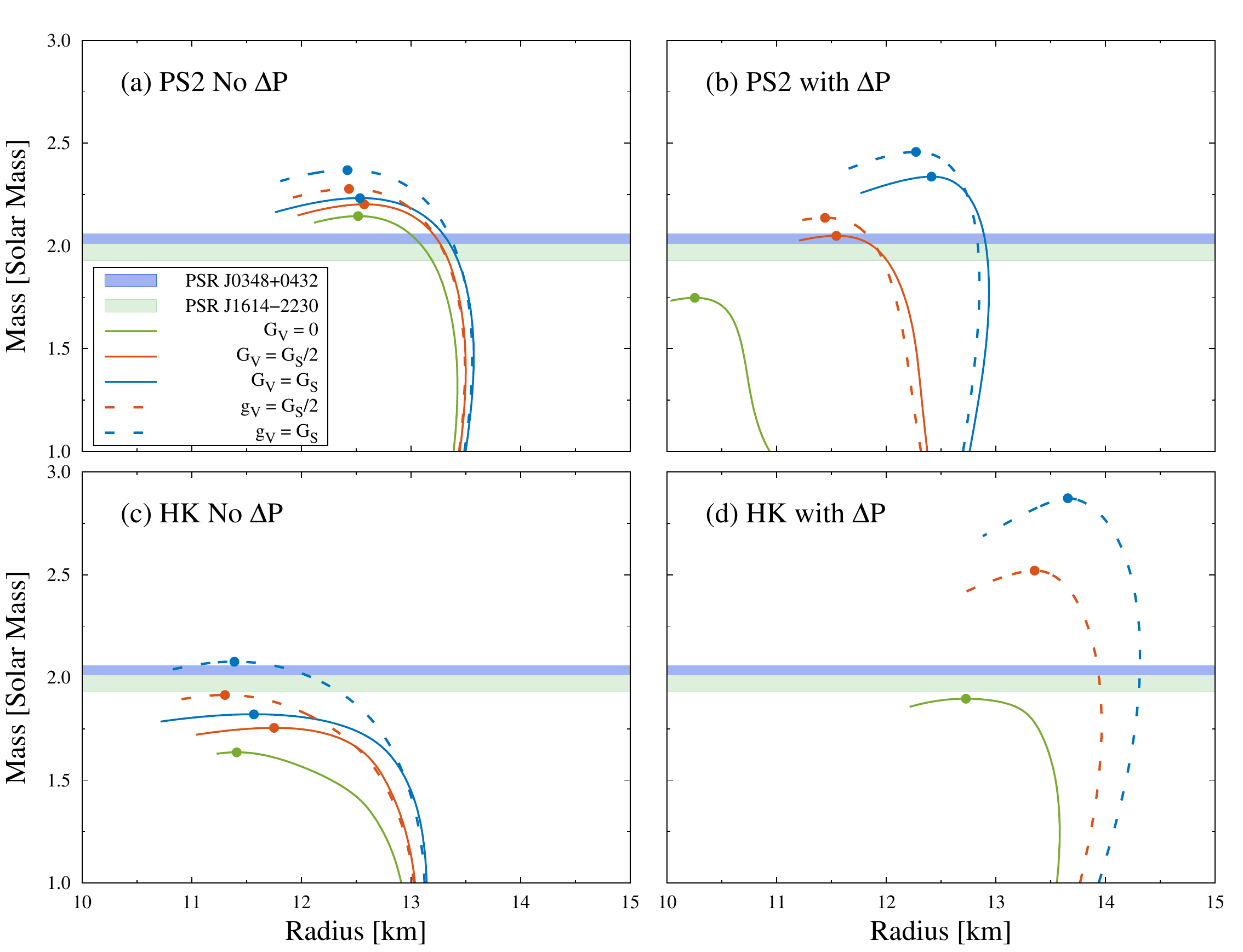} 
\caption{Neutron star mass as a function of radius. For plots (a,b), the interpolation is between the ``Standard" or baseline scenario of the HF-QMC model and the proper time regularised PS2 model with flavour dependent (solid) and independent (dashed) vector interactions. Similarly for plots (c,d), but with the three momentum regularised model with flavour dependent (solid) and independent (dashed) vector interactions. Plots (a,c) do not include the thermodynamic correction $\Delta P$, whereas plots (b,d) include the correction for thermodynamic consistency. The crossover region is chosen to be $\left( \bar{\rho},\Gamma\right) =\left( 3\rho_{0},\rho_{0}\right)$. Specific curves for all plots are indicated in the key of plot (a). }
\label{fig:MVsR_Multi}
\end{figure*}

Figure~\ref{fig:PressVsEnDen_Multi} shows pressure with and without the thermodynamic correction as a function of energy density. 
As can be seen by comparing curves with and without the thermodynamic correction, the interpolated EoS is significantly affected by $\Delta P$ in the transition region. Without the thermodynamic correction, the transitions between the hadronic and all quark EoS occur smoothly without violating the constraints of thermodynamic stability and causality  as would be expected from monotonic functions like those expressed in Eq.~(\ref{eq:fpm}). Also, away from the transition region $(\bar{\rho},\Gamma)=(3\rho_{0},\rho_{0})$ the interpolated EoS are almost equivalent to the un-interpolated hadronic and quark EoS. However, in plot (d) of Fig.~\ref{fig:PressVsEnDen_Multi}, the correction to the pressure is so significant that the resulting EoS becomes unstable for HK models with $G_{\rm V} = G_{\rm S}/2$ and $G_{\rm V} = G_{\rm S}$. The greater the separation in the $\epsilon$--$\rho$ plane between the EoS of the two phases, the larger the correction $\Delta P$, leading to a more significant change in the pressure.

The speed of sound squared in matter is shown in Fig.~\ref{fig:Cs2VsEnDen_Multi} as a function of energy density. Interpolations between the hadronic EoS and quark models with both types of vector interaction are included in Fig.~\ref{fig:Cs2VsEnDen_Multi}.
On comparing with Fig.~\ref{fig:PressVsEnDen_Multi}, it can be seen that as an EoS softens, sound slows down and as the EoS
stiffens, sound speeds up. The speed of sound is a very good measure of the stiffness of an EoS. Without the thermodynamic correction the interpolated EoS with PS2 models are generally stiffer than those with HK models, particularly at low and intermediate density. It is also clear that the flavour independent vector interaction produces a stiffer EoS and hence faster sound than models incorporating the flavour dependent vector interaction. On inclusion of the thermodynamic correction, however, interpolated EoS with HK models incorporating a vector interaction  are stiffer than their PS2 counterparts. 
The speed of sound is enhanced in the transition region for both PS2 and HK models. However, it is not as significant for the PS2 models. 
The EoS used in plotting the curves in Fig.~\ref{fig:Cs2VsEnDen_Multi} remain causal and
most interpolated EoS examined in the tables also remain causal. However, there were a few exceptions. Those that did not meet the stability and causality requirements are indicated by asterisks ($\ast$) in Tables~\ref{table:3rho0} and \ref{table:VaryRhoBar}. In plot (d) of Fig.~\ref{fig:Cs2VsEnDen_Multi}, the speed of sound becomes imaginary for HK models with $G_{\rm V} = G_{\rm S}/2$ and $G_{\rm V} = G_{\rm S}$, once again indicating an unstable EoS .

If an interpolated EoS was found to be unstable or to violate causality, that does not mean an interpolation between those particular hadronic and quark models is not possible in general. It simply means it is not possible to construct a consistent EoS in our current interpolation scheme. Equations of state that do not meet our requirements of stability and causality with the chosen interpolation scheme are simply discarded. More detailed investigations on the dependence of the interpolation scheme are beyond the scope of this paper and are left for future work. 

\subsection{Hybrid stars}
Interpolated EoS which satisfied the two constraints were used as input to the TOV equations~\footnote{The BPS EoS was attached at low density as in Sec.~\ref{sec:gbem}}. The resulting mass-radius ($M$--$R$) relations for several interpolations, with and without the thermodynamic correction, are shown in Fig.~\ref{fig:MVsR_Multi}. All curves shown in Fig.~\ref{fig:MVsR_Multi} use the ``Standard" or baseline scenario of the HF-QMC model, so all variations are a result of changes in the quark model and the thermodynamic correction $\Delta P$. Without $\Delta P$, interpolations with PS2 models are shown to produce massive hybrid stars, compatible with the observations of Demorest {\it et al.}~\cite{Demorest:2010bx} and Antoniadis {\it et al.}~\cite{Antoniadis:2013pzd}, even in the absence of a vector interaction. However, for HK models only when $g_{\rm V} = G_{\rm S}$ is a sufficiently massive hybrid star, compatible with observations, actually  produced.
 The only difference between these interpolated EoS is the quark model. The prediction of more massive  hybrid stars is a result of the stiffer PS2 quark EoS at low and intermediate density. The softer HK models predict radii about $0.5$~km smaller than PS2 models, but for both models increasing the vector coupling increases the radius only slightly.

Including $\Delta P$ has a significant impact on the $M$--$R$ relationships for all models. Considering the interpolated EoS with the PS2  quark models, those  with $G_{\rm V} = 0$ and $g_{\rm V} =G_{\rm V} = G_{\rm S}/2$ predict smaller maximum masses, whereas the $g_{\rm V} =G_{\rm V} = G_{\rm S}$ models produce more massive hybrid stars. In the absence of a vector interaction the interpolated EoS with the PS2 model no longer satisfies the constraints set by the observations of massive stars. More noticeable, however,  is the separation of curves in terms of radii.  The softening at the start of the transition region coming from the correction $\Delta P$ significantly reduces the radius, particularly for $G_{\rm V} = 0$. The other cases also yield smaller radii than when $\Delta P$ is ignored. As for the interpolated EoS utilising the HK models, the maximum masses are considerably larger and the radii are bigger because $\Delta P$ is always positive and larger in magnitude. The model with $G_{\rm V} = 0$ still does not meet the astrophysical constraints.

A summary of maximum mass configurations is presented in Table~\ref{table:3rho0}. To show the dependence of the interpolated EoS on the hadronic EoS we also included interpolations between the overly stiff variation of the HF-QMC model, where the cut-off used in the Fock terms is increased from $\Lambda =0.9$~GeV to $\Lambda = 1.3$~GeV. From this table, it can be inferred that while the properties of the maximum mass configurations of hybrid stars are affected by the hadronic model, they are much more sensitive to the quark model.

Table~\ref{table:VaryRhoBar} summarises hybrid star properties under variation of the transition region, $\bar{\rho}\in \left\lbrace 3\rho_{0},4\rho_{0},5\rho_{0}\right\rbrace $. As would naively be expected, pushing the centre of the transition region to higher density tends to produce a less stiff EoS for the majority of the interpolations which naturally translates to smaller maximum masses for hybrid stars. On delaying the transition to higher densities, it was found to be more difficult to construct a consistent EoS between the chosen hadronic and quark models. This was conspicuously evident for the HK models, partly owning to the greater separation in the $\epsilon$--$\rho$ plane of the hadronic and quark model curves, leading to a larger thermodynamic correction $\Delta P$.

 \begin{table*}%[H]% add [H] placement to break table across pages

 \begin{ruledtabular}
 \begin{tabular}{cccccccccc}
Hadronic & Quark & $G_{\rm V}$ & Vector  & \multicolumn{2}{c}{$M_{\rm max}$ [$M_{\odot}$] } & \multicolumn{2}{c}{$R$ [km]}   & \multicolumn{2}{c}{$\rho^{\rm max}_{c}$ [$\rho_{0}$]} \\
Model & Model & &Int. & No $\Delta P$ & $\Delta P$ & No $\Delta P$ & $\Delta P$ & No $\Delta P$ & $\Delta P$  \\
\hline
Standard &  PS2 & 0 &  - & 2.15 &1.75 & 12.52&  10.25 & 5.50 & 7.54 \\
Standard &  PS2 & $G_{\rm S}/2$& dep. & 2.20 &2.05 & 12.57 & 11.54 & 5.18& 5.90 \\
Standard &  PS2 & $G_{\rm S}/2$ & indep. & 2.28& 2.14& 12.44 & 11.44 & 5.28& 5.94 \\
Standard &  PS2 & $G_{\rm S}$ & dep.&2.23 & 2.34 & 12.54 & 12.41& 5.00 & 4.98 \\
Standard &  PS2 & $G_{\rm S}$ & indep.& 2.37 & 2.46& 12.42& 12.27 & 5.02& 5.02 \\
\hline
Standard &  HK & 0 & - &1.64 & 1.90 & 11.41 & 12.73 & 6.32 & 4.90\\
Standard &  HK & $G_{\rm S}/2$& dep.  & 1.76 &$\ast$& 11.75 &$\ast$& 5.40 &$\ast$ \\
Standard &  HK & $G_{\rm S}/2$ & indep. & 1.92 & 2.52& 11.30 & 13.35 & 5.92 & 4.36  \\
Standard &  HK & $G_{\rm S}$ & dep.& 1.82 &$\ast$& 11.57&$\ast$& 5.38&$\ast$ \\
Standard &  HK & $G_{\rm S}$ & indep.& 2.01 & 2.87& 11.39 & 13.66 & 5.40 & 4.08 \\
\hline
$\Lambda = 1.3$ &  PS2 & 0 & - &2.19 & 1.73 & 12.74  & 10.24 & 5.34 & 7.68\\
$\Lambda = 1.3$ &  PS2 & $G_{\rm S}/2$& dep. &  2.25  & 2.03 & 12.79 & 11.57 & 5.02 & 5.94 \\
$\Lambda = 1.3$ &  PS2 & $G_{\rm S}/2$ & indep. & 2.32 & 2.12 & 12.64 & 11.46& 5.14 & 5.98  \\
$\Lambda = 1.3$ &  PS2 & $G_{\rm S}$ & dep.& 2.28 & 2.32 & 12.75 & 12.48 & 4.86 & 4.96 \\
$\Lambda = 1.3$ &  PS2 & $G_{\rm S}$ & indep.& 2.40 & 2.44 &  12.61 & 12.31 & 4.90 & 5.02  \\
\hline
$\Lambda = 1.3$ &  HK & 0 &  - &1.69 & 1.89 & 11.73 & 12.95& 5.96 & 4.66 \\
$\Lambda = 1.3$ &  HK & $G_{\rm S}/2$& dep. & 1.81 &$\ast$&  12.12 &$\ast$&  5.04 &$\ast$ \\
$\Lambda = 1.3$ &  HK & $G_{\rm S}/2$ & indep. & 1.95 & 2.52 & 11.56 & 13.47 & 5.70 & 4.28 \\
$\Lambda = 1.3$ &  HK & $G_{\rm S}$ & dep.& 1.87 &$\ast$& 11.97  &$\ast$&  5.00 &$\ast$\\
$\Lambda = 1.3$ &  HK & $G_{\rm S}$ & indep.&  2.11  &  2.87 & 11.62  & 13.73  & 5.24  & 4.04  \\
\end{tabular}
 \end{ruledtabular}
 \caption{Hybrid star properties in the percolation picture. The crossover region is chosen to be $\left( \bar{\rho},\Gamma\right) =\left( 3\rho_{0},\rho_{0}\right)$. An asterisk ($\ast$) indicates that a consistent EoS could not be constructed between that variation of hadronic and quark models with the chosen interpolation method. Tabulated hybrid star properties are maximum stellar mass, stellar radius and corresponding central density.}
\label{table:3rho0}
 \end{table*}
%}

{\squeezetable
\begin{table}%[H] %add [H] placement to break table across pages
 \begin{ruledtabular}
 \begin{tabular}{ccccccccc}
 Quark & $G_{\rm V}$ & $\bar{\rho}$   & \multicolumn{2}{c}{$M_{\rm max}$ [$M_{\odot}$] } & \multicolumn{2}{c}{$R$ [km]}   & \multicolumn{2}{c}{$\rho^{\rm max}_{c}$ [$\rho_{0}$]} \\
 Model &  && No $\Delta P$ & $\Delta P$ & No $\Delta P$ & $\Delta P$ & No $\Delta P$ & $\Delta P$  \\
\hline
  PS2 & 0 &$3\rho_{0}$&  2.15 & 1.75& 12.52 & 10.25  &5.5 & 7.54 \\
  PS2 & 0  &$4\rho_{0}$& 1.98 &$\ast$&  11.84 &$\ast$&  6.18  &$\ast$ \\
  PS2 & 0 &$5\rho_{0}$&  1.87 &$\ast$& 11.54 &$\ast$ & 6.56 & $\ast$ \\

  PS2 & $G_{\rm S}/2$&$3\rho_{0}$& 2.20 & 2.05 & 12.57 & 11.54 & 5.18 & 5.90 \\
  PS2 & $G_{\rm S}/2$&$4\rho_{0}$& 2.01 & 1.99 & 11.74 & 11.43 & 5.96  & 6.30\\
  PS2 & $G_{\rm S}/2$&$5\rho_{0}$& 1.89 & 1.96 & 11.36 & 11.20 & 6.42  & 6.78\\

  PS2 & $G_{\rm S}$  &$3\rho_{0}$&  2.23  &  2.34  & 12.54 &  12.41  & 5.00 & 4.98 \\
  PS2 & $G_{\rm S}$  &$4\rho_{0}$&  2.03  &  2.28  &  11.61  &  11.85  &  5.82  & 5.72 \\
  PS2 & $G_{\rm S}$  &$5\rho_{0}$&  1.91  &$\ast$&  11.17  &$\ast$&  6.36   &$\ast$ \\

\hline
  HK & 0 & $3\rho_{0}$ &1.64  &  1.90  &  11.41  &  12.73  & 6.32  &  4.90 \\
  HK & 0 & $4\rho_{0}$ &  1.74  &  1.92  &  11.91  &  12.24  &  5.70  &  5.50  \\
  HK & 0 & $5\rho_{0}$ &  1.78  &$\ast$&  12.00 &$\ast$&  5.60 & $\ast$\\

  HK & $G_{\rm S}/2$ & $3\rho_{0}$ &  1.76  &$\ast$&  11.75  &$\ast$&  5.40 & $\ast$\\
  HK & $G_{\rm S}/2$ & $4\rho_{0}$ &  1.78  &$\ast$&  11.92  &$\ast$& 5.26  & $\ast$\\
  HK & $G_{\rm S}/2$ & $5\rho_{0}$ &  1.79  &$\ast$&  11.93  &$\ast$&  5.38   & $\ast$\\

  HK & $G_{\rm S}$ &$3\rho_{0}$&  1.82  &$\ast$&  11.57  &$\ast$&  5.38   & $\ast$\\
  HK & $G_{\rm S}$ &$4\rho_{0}$&  1.82  &$\ast$&  11.59  &$\ast$&   5.38  & $\ast$\\
  HK & $G_{\rm S}$ &$5\rho_{0}$&  1.81  &$\ast$&  11.61  &$\ast$&  5.50   & $\ast$\\

 \end{tabular}
 \end{ruledtabular}
 \caption{Hybrid star properties in the percolation picture under variation of $\bar{\rho}\in \left\lbrace 3\rho_{0},4\rho_{0},5\rho_{0}\right\rbrace $. The ``Standard" or baseline scenario is used for the hadronic model and the flavour dependent vector interaction is used in each of the quark models. An asterisk ($\ast$) indicates that a consistent EoS could not be constructed between that variation of hadronic and quark models with the chosen interpolation method. Tabulated hybrid star properties are maximum stellar mass, stellar radius and corresponding central density.}
\label{table:VaryRhoBar}
 \end{table}
}

\section{Summary}
\label{sec:summary}

%%%%
We began by introducing the Hartree-Fock QMC and NJL models. The particle content and EoS of hadronic and quark matter were presented
in Secs.~\ref{sec:gbem} and \ref{sec:quark}, respectively. The numerical results of our hadronic calculation were presented in detail in Ref.~\cite{PhysRevC.89.065801}, so we place greater emphasis on the quark matter results and the construction of a crossover EoS.

The results of the proper time regularised NJL model developed in this work were compared to the three momentum regularised NJL model with t' Hooft determinant term. The proper time NJL parameter set PS2 was preferred for modelling high density matter over PS1, because of the behaviour of the constituent quark mass  as a function of chemical potential.  For quark matter in beta equilibrium, the PS2 model produced overall qualitatively similar results to the HK model, despite the different behaviour of the quark masses. However, the PS2 model produced slightly higher pressure, particularly at low density, as compared with the HK model. 
%%%%%%

Following this we discussed phase transitions from hadronic matter to quark matter, with emphasis on the conventional first order treatments via the Maxwell and Gibbs constructions. How they are implemented, their properties and their shortcomings were highlighted. We then discussed modeling the transition as a smooth crossover, a possibility which has recently been given much consideration in the literature~\cite{Asakawa1997,Masuda:2012kf,Masuda:2012ed,PhysRevC.90.045801,AlvarezCastillo2014,Kojo2015}. Motivation for such a transition was discussed and one method for implementing this kind of transition was presented. The numerical results for the faux crossover construction between the HF-QMC and NJL models developed in earlier sections were then presented and discussed.

The interpolation transitions for the ``Standard" and the ``$\Lambda = 1.3$~GeV" hadronic matter EoS scenarios from Sec.~\ref{sec:hadronic} were shown.  It was concluded that the hadronic model has only a small effect on the maximum neutron star mass and it is mostly dependent on the quark model.

The effect of this interpolation method on the EoS of dense matter is shown in Fig.~\ref{fig:PressVsEnDen_Multi}. At low and high density it can be seen to approach the assumed asymptotic limits, i.e. the hadronic and quark EoS, but in the intermediate region the pressure can be somewhat weakened or enhanced depending on $\Delta P$. It produces a decrease in the pressure at the beginning of the transition region and an increase towards the end for PS2 models and the opposite behaviour for HK models. This clearly comes from the density dependence of the difference between the energy densities of the two EoS (see Eq.~(\ref{eq:DeltaP})) and is therefore dependent on the two EoS between which we are interpolating. Moreover, it suggests that the pressure in the transition region can potentially be outside the limits set by the hadronic and quark EoS and may possibly have inflection points leading to an instability or violation of causality.

The main conclusion of this work is that the observations of large neutron stars can certainly be explained within such a construction using the HF-QMC and NJL models, provided the quark model is sufficiently stiff and the transition occurs at low density, $\bar{\rho}\sim 3\rho_{0}$. This is in agreement with other recent works using similar and alternative methods to phenomenologically implement a faux crossover between point-like hadronic and quark models. 

Another important conclusion is that the correction $\Delta P$ (Eq.~(\ref{eq:DeltaP})) has a considerable influence on the interpolated EoS. As it arises from calculating the pressure from the phenomenologically interpolated energy density, its meaning is somewhat ambiguous. It is required for thermodynamic consistency, but on one hand it may merely be an artefact of the chosen interpolation scheme. On the other hand, it could be associated with non-perturbative physical effects in the transition region. More in depth work is certainly needed to understand the validity and importance of this term and the overall dependence on the interpolation scheme. Further insight can only come from more detailed analysis of QCD thermodynamics above $2\rho_{0}$. Future heavy ion collision experiments probing this region will certainly play an important role.

It must be stressed that the hybrid EoS developed here hinge on the assumption that the deconfinement transition is a crossover and that we can smoothly interpolate between hadronic and quark models, essentially parametrising our ignorance of the intermediate transition region. Determining whether it is indeed a crossover or not may be possible with upcoming heavy ion collision experiments probing even higher densities and greater asymmetries. If this were found to be a valid assumption, it would offer a possible resolution to the problem of exotic degrees of freedom and massive compact stars, although, more work would be needed to understand the interpolation dependence and the physical meaning behind corrections such as $\Delta P$.
%

%\appendix
%\section{}

\begin{acknowledgments}
This work was supported by the University of Adelaide and the Australian
Research Council through grants FL0992247  and DP151103101 (AWT).
\end{acknowledgments}

\bibliography{QMC_NJL_QM_paper}

\end{document}